\documentclass{ws-rv9x6}

\usepackage{hyperref}
\usepackage{multirow}
\usepackage{arydshln}

\catcode`\@=11
\def\@listI{\leftmargin\leftmargini
            \listparindent\itemindent
            \parsep \z@\labelsep.5em
            \topsep 0\p@ \@plus0\p@
            \itemsep0\p@}
\catcode`@=12

\def\lsim{\mathrel{\mathpalette\vereq<}}
\def\gsim{\mathrel{\mathpalette\vereq>}}
\def\vereq#1#2{\lower3.5pt\vbox{\baselineskip1.5pt \lineskip1.5pt
\ialign{$#1\hfill##\hfil$\crcr#2\crcr\sim\crcr}}}

\newcommand{\nn}{\nonumber}
\newcommand{\beq}{\begin{equation}}
\newcommand{\eeq}{\end{equation}}
\newcommand{\beqa}{\begin{eqnarray}}
\newcommand{\eeqa}{\end{eqnarray}}
\newcommand{\TeV}{{\rm TeV}}
\newcommand{\GeV}{{\rm GeV}}

\def\ov{\overline}
\def\rhobar{\bar\rho}
\def\etabar{\bar\eta}
\def\lqcd{\Lambda_{\rm QCD}}
\newcommand{\ds}{\displaystyle}
\def\d{{\rm d}}

\newcommand{\Bbar}{\,\overline{\!B}{}}
\newcommand{\Dbar}{\,\overline{\!D}{}}
\newcommand{\Kbar}{\,\overline{\!K}{}}
\def\B0bar{\Bbar{}^0}
\def\D0bar{\Dbar{}^0}
\def\K0bar{\Kbar{}^0}

\def\psla{p\hspace{-4.5pt}\slash}
\def\ksla{k\hspace{-5pt}\slash}
\def\vsla{v\hspace{-4.75pt}\slash}
\def\Dsla{D\hspace{-6.5pt}\slash\,}
\def\doteq{\rlap{\raise 3pt\hbox{$\;\cdot$}}{=}}

\begin{document}

\chapter*{TASI Lectures on Flavor Physics}

\author{Zoltan Ligeti}

\address{Ernest Orlando Lawrence Berkeley National Laboratory,\\
University of California, Berkeley, CA 94720\\
$^*$E-mail: ligeti@berkeley.edu}

\begin{abstract}

These notes overlap with lectures given at the TASI summer schools in 2014 and
2011, as well as at the European School of High Energy Physics in 2013.  This is
primarily an attempt at transcribing my hand-written notes, with emphasis on
topics and ideas discussed in the lectures.  It is not a comprehensive
introduction or review of the field, nor does it include a complete list of
references.  I hope, however, that some may find it useful to better
understand the reasons for excitement about recent progress and future
opportunities in flavor physics.

\end{abstract}

\body

\section*{Preface}

There are many books and reviews on flavor physics (e.g.,
Refs.~\cite{Branco:1999fs, Manohar:2000dt, Ligeti:2002wt, Ligeti:2003fi,
Nir:2005js, Hocker:2006xb, Grossman:2009dw, Grinstein:2015nya, PDG}).
The main points I would like to explain in these lectures are:
\begin{itemize}

\item $CP$ violation and flavor-changing neutral currents (FCNC) are sensitive
probes of short-distance physics, both in the standard model (SM) and  in beyond
standard model (BSM) scenarios.

\item The data taught us a lot about not directly seen physics in the past, and
are likely crucial to understand LHC new physics (NP)~signals.

\item In most FCNC processes BSM/SM $\sim{\cal O}(20\%)$ is still allowed today,
the sensitivity will improve to the few percent level in the future.

\item Measurements are sensitive to very high scales, and might find
unambiguous signals of BSM physics, even outside the LHC reach.

\item There is a healthy and fun interplay of theoretical and experimental
progress, with many open questions and important problems.

\end{itemize}
Flavor physics is interesting because there is a lot we do not understand yet. 
The ``standard model flavor puzzle" refers to our lack of understanding of
why and how the 6 quark and 6 lepton flavors differ, why masses and quark
mixing are hierarchical, but lepton mixing is not.  The ``new physics flavor
puzzle" is the tension between the relatively low scale required to solve the
fine tuning problem (also suggested by the WIMP paradigm), and the high scale
that is seemingly required to suppress the non-SM contributions to flavor
changing processes.  If there is NP at the TeV scale, we need to understand why
and in what way its flavor structure is non-generic.

The key questions and prospects that make the future
interesting are~\cite{Grossman:2009dw}
\begin{itemize}

\item What is the achievable experimental precision?

The LHCb, Belle~II, NA62, KOTO, $\mu\to e\gamma$, $\mu 2e$, etc., experiments 
will improve the sensitivity in many modes by orders of magnitude.

\item What are the theoretical uncertainties?

In many key measurements, the theory uncertainty is well below future
experimental sensitivity; while in some cases theoretical improvements are
needed (so you can make an impact!).

\item How large deviations from SM can we expect due to TeV-scale NP?

New physics with generic flavor structure is ruled out; observable effects
near current bounds are possible, many models predict some.

\item What will the measurements teach us?

In all scenarios there is complementarity with high-$p_T$ measurements, and
synergy in understanding the structure of any NP seen.

\end{itemize}

Another simple way to get a sense of (a lower bound on) the next 10--15 years of
$B$ physics progress is to consider the expected increase in data,
\beq
\frac{\mbox{(LHCb upgrade)}}{\mbox{(LHCb $1\,{\rm fb}^{-1}$)}} \sim
\frac{\mbox{(Belle II data set)}}{\mbox{(Belle data set)}} \sim
\frac{\mbox{(2009 BaBar data set)}}{\mbox{(1999 CLEO data set)}} \sim 50\,.\nn
\eeq
This will yield a $\sqrt[4]{50} \sim 2.5$ increase in sensitivity to higher mass
scales, even just by redoing existing measurements.  More data has always
motivated new theory ideas, yielding even faster progress.  This is a comparable
increase in reach as going from LHC7--8 $\to$ LHC13--14.

\paragraph{Outline}

The topics these lectures will cover include a brief introduction to flavor
physics in the SM, testing the flavor structure in neutral meson mixing and $CP$
violation, and examples of how to get theoretically clean information on
short-distance physics.  After a glimpse at the ingredients of the SM CKM fit,
we discuss how sizable new physics contributions are still allowed in neutral
meson mixing, and how this will improve in the future.  Then we explain some
implications of the heavy quark limit, tidbits of heavy quark symmetry, the
operator product expansion and inclusive decays, to try to give an impression of
what makes some hadronic physics tractable.  The last lecture discusses some
topics in TeV-scale flavor physics, top quark physics, Higgs flavor physics,
bits of the interplay between searches for supersymmetry and flavor, and
comments on minimal flavor violation.  Some questions one may enjoy thinking
about are in the footnotes.

\section{Introduction to Flavor Physics and $CP$ Violation}
\addcontentsline{toc}{}{Introduction to Flavor Physics and $CP$ Violation}

Most of the experimentally observed particle physics phenomena are consistent
with the standard model (SM).  Evidence that the minimal SM is incomplete comes
from the lack of a dark matter candidate, the baryon asymmetry of the Universe,
its accelerating expansion, and nonzero neutrino masses.  The baryon asymmetry
and neutrino mixing are certainly connected to $CP$ violation and flavor
physics, and so may be dark matter.  The hierarchy problem and seeking to
identify the particle nature of dark matter strongly motivate TeV-scale new
physics.

Studying flavor physics and $CP$ violation provides a rich program to probe the
SM and search for NP, with sensitivity to the 1\,--\,$10^5\,$TeV scales,
depending on details of the models.  As we shall see, the sensitivity to BSM
contributions to the dimension-6 four-quark operators mediating $K$, $D$, $B_d$,
and $B_s$ mixing, when parametrized by coefficients $1/\Lambda^2$, corresponds
to scales $\Lambda \sim 10^2 - 10^5\,\TeV$ (see Table~\ref{tab:mixingscales} and
the related discussion below).

Understanding the origin of this sensitivity and how it can be improved,
requires going into the details of a variety of flavor physics measurements.

\paragraph{Baryon asymmetry requires $CP$ violation beyond SM}

The baryon asymmetry of the Universe is the measurement of
\beq
\frac{n_B - n_{\bar B}}s \approx 10^{-10}\,,
\eeq
where $n_B$ ($n_{\bar B}$) is the number density of (anti-)baryons and $s$ is
the entropy density.
This means that $10^{-6}$ seconds after the Big Bang, when the temperature was
$T>1\,\GeV$, and quarks and antiquarks were in thermal equilibrium, there was a
corresponding asymmetry between quarks and antiquarks. Sakharov pointed
out~\cite{Sakharov:1967dj} that for a theory to generate such an asymmetry in
the course of its evolution from a hot Big Bang (assuming inflation washed out
any possible prior asymmetry), it must contain:
\begin{enumerate}
\item baryon number violating interactions;
\item $C$ and $CP$ violation;
\item deviation from thermal equilibrium.
\end{enumerate}
Interestingly, the SM contains 1--2--3, but (i) $CP$ violation is too
small, and (ii) the deviation from thermal equilibrium is too small at the
electroweak phase transition.  The SM expectation is many orders of magnitude
below the observation, due to the suppression of $CP$ violation by
\beq
\big[\Pi_{u_i \neq u_j} (m_{u_i}^2 - m_{u_j}^2)\big]
  \big[\Pi_{d_i \neq d_j} (m_{d_i}^2 - m_{d_j}^2)\big] / m_W^{12}\,,
\eeq
and $m_W$ indicates a typical weak interaction scale here.\footnote{Why is this
suppression a product of all up and down quark mass differences, while fewer
factors of mass splittings suppress $CP$ violation in hadron decays and meson
mixings?}

Therefore, $CP$ violation beyond the SM must exist.  While this argument does
not tell us the scale of the corresponding new physics, it motivates searching
for new sources of $CP$ violation.  (It may occur only in flavor-diagonal
processes, such as EDMs, or only in the lepton sector, as in leptogenesis.)  In
any case, we want to understand the microscopic origin of $CP$ violation, and
how precisely we can test those $CP$-violating processes that we can measure.

Equally important is that almost all TeV-scale new physics models contain new
sources of $CP$ violation.  Baryogenesis at the electroweak scale may still be
viable, and the LHC will probe the remaining parameter space.

\paragraph{The SM and flavor}

The SM is defined by the gauge interactions,
\beq
SU(3)_c \times SU(2)_L \times U(1)_Y\,,
\eeq
the particle content, i.e., three generations of the fermion representations,
\beq\label{reps}
Q_L(3,2)_{1/6},\quad u_R(3,1)_{2/3},\quad d_R(3,1)_{-1/3},\quad
L_L(1,2)_{-1/2},\quad \ell_R(1,1)_{-1}\,,
\eeq
and electroweak symmetry breaking.  A condensate $\langle \phi \rangle =
\Big(\begin{matrix} 0\\ v/\sqrt2 \end{matrix}\Big)$ breaks $SU(2)_L \times
U(1)_Y \to  U(1)_{\rm EM}$, the dynamics of which we now know is well
approximated by a seemingly elementary SM-like scalar Higgs field.

The kinetic terms in the SM Lagrangian are
\beq\label{Lkin}
{\cal L}_{\rm kin} = 
  -\frac14 \sum_{\rm groups} (F_{\mu\nu}^a)^2 
  + \sum_{\rm rep's} \ov\psi\, i \Dsla\, \psi\,.
\eeq
These are always $CP$ conserving, as long as we neglect a possible $F\widetilde
F$ term.  The ``strong $CP$ problem"~\cite{Dine:2000cj} is the issue of why the
coefficient of the $F\widetilde F$ term for QCD is tiny.  Its solution is an
open question; however, we know that it is negligible for flavor-changing
processes.  The Higgs terms,
\beq\label{Lhiggs}
{\cal L}_{\rm Higgs} = |D_\mu \phi|^2 + \mu^2 \phi^\dagger\phi
  - \lambda (\phi^\dagger\phi)^2\,,
\eeq
are $CP$ conserving in the SM, but can be $CP$ violating with an extended Higgs
sector (already with two Higgs doublets; three are needed if natural flavor
conservation is imposed~\cite{Glashow:1976nt}).  Finally, the Yukawa
couplings are,
\beq\label{Lyuk}
{\cal L}_Y = - Y_{ij}^d\, \ov{Q_{Li}^I}\, \phi\, d_{Rj}^I
- Y_{ij}^u\, \ov{Q_{Li}^I}\, \widetilde\phi\, u_{Rj}^I
- Y_{ij}^\ell\, \ov{L_{Li}^I}\, \phi\, \ell_{Rj}^I + {\rm h.c.}
\eeq
The $Y^{ij}_{u,d}$ are $3\times3$ complex matrices, $i,j$ are generation
indices, $\widetilde\phi = i\sigma_2\phi^*$.

After electroweak symmetry breaking, Eq.~(\ref{Lyuk}) gives quark mass terms,
\beqa\label{mass}
{\cal L}_{\rm mass} &=&{} - \ov{d_{Li}^I}\, (M_d)_{ij}\, d_{Rj}^I
  - \ov{u_{Li}^I}\, (M_u)_{ij}\, u_{Rj}^I + {\rm h.c.} \nn\\
&=&{} - \big(\ov{d_{L}^I} V_{dL}^\dagger\big)
  \big( V_{dL} M_d V_{dR}^\dagger \big) \big(V_{dR}\, d_{R}^I\big) \nn\\
&&{} - \big(\ov{u_{L}^I} V_{uL}^\dagger\big)
  \big( V_{uL} M_u V_{uR}^\dagger\big) \big(V_{uR}\, u_{R}^I\big)
  + {\rm h.c.},
\eeqa
where $M_f = (v/\sqrt2)\, Y^f$.  The last two lines show the diagonalization of
the mass matrices necessary to obtain the physical mass eigenstates,
\beq
M_f^{\rm diag} \equiv V_{fL}\, M_f\, V_{fR}^\dagger\,, \qquad
f_{Li} \equiv V_{fL}^{ij}\, f_{Lj}^I\,, \qquad
f_{Ri} \equiv V_{fR}^{ij}\, f_{Rj}^I\,,
\eeq
where $f=u,d$ denote up- and down-type quarks.  The diagonalization is
different for $u_{Li}$ and $d_{Li}$, which are in the same $SU(2)_L$ doublet,
\beq\label{mismatch}
\bigg(\begin{matrix} u_{Li}^I\\ d_{Li}^I\end{matrix}\bigg) 
  = (V_{uL}^\dagger)_{ij}\, \bigg( \begin{matrix} 
  u_{Lj}\\ (V_{uL}V_{dL}^\dagger)_{jk}\, d_{Lk}\end{matrix}\bigg) \,.
\eeq
The ``misalignment" between these two transformations,
\beq\label{CKMdef}
V_{\rm CKM} \equiv V_{uL} V_{dL}^\dagger \,,
\eeq
is the Cabibbo-Kobayashi-Maskawa (CKM) quark mixing matrix.  By virtue of
Eq.~(\ref{CKMdef}), it is unitary.

Eq.~(\ref{mismatch}) shows that the charged current weak interactions, which
arise from the $\ov\psi\, i \Dsla\, \psi$ terms in Eq.~(\ref{Lkin}), become
non-diagonal in the mass basis
\beq
- {g\over 2}\, \ov{Q_{Li}^I}\, \gamma^\mu W_\mu^a \tau^a Q_{Li}^I 
  + {\rm h.c.} \ \Rightarrow \ 
- {g\over \sqrt2}\ \big(\ov{u_{L}},\ \ov{c_{L}},\, \ov{t_{L}}\big)
  \gamma^\mu W_\mu^+\, V_{\rm CKM}
  \left(\begin{matrix}d_L \\ s_L \cr b_L\end{matrix}\right) + {\rm h.c.},
\eeq
where $W_\mu^\pm = (W^1_\mu \mp W^2_\mu) / \sqrt2$.  Thus, charged-current weak
interactions change flavor, and this is the only flavor-changing interaction in
the SM.

In the absence of Yukawa couplings, the SM has a global $[U(3)]^5$ symmetry
($[U(3)]^3$ in the quark and $[U(3)]^2$ in the lepton sector), rotating the 3
generations of the 5 fields in Eq.~(\ref{reps}).  This is broken by the Yukawa
interactions in Eq.~(\ref{Lyuk}).  In the quark sector the breaking is
\beq\label{quarkglobal}
U(3)_Q \times U(3)_u \times U(3)_d \, \to\,  U(1)_B\,,
\eeq
In the lepton sector, we do not yet know if $U(3)_L \times U(3)_\ell$ is fully
broken.

\paragraph{Flavor and $CP$ violation in the SM}

Since the $Z$ couples flavor diagonally,\footnote{Show that there are no
tree-level flavor-changing $Z$ couplings in the SM.  What if, besides doublets,
there were a left-handed $SU(2)$ singlet quark field as well?} there are no
tree-level flavor-changing neutral currents, such as $K_L \to \mu^+\mu^-$.  This
led GIM~\cite{Glashow:1970gm} to predict the existence of the charm quark. 
Similarly, $K^0$\,--\,$\K0bar$ mixing vanishes at tree-level, which allowed the
prediction of $m_c$~\cite{Vainshtein:1973md, Gaillard:1974hs} before the
discovery of the charm quark.  In the previous examples, because of the
unitarity of the CKM matrix,
\beq\label{Ktriangle}
V_{ud}\, V_{us}^* + V_{cd}\, V_{cs}^* + V_{td}\, V_{ts}^* = 0\,.
\eeq
Expanding the loop functions, e.g., in a FCNC kaon decay amplitude,
\beq
V_{ud}\, V_{us}^*\, f(m_u)+ V_{cd}\, V_{cs}^*\, f(m_c)+ V_{td}\,
V_{ts}^*\, f(m_t)\,,
\eeq
the result is always proportional to the up-quark mass-squared differences,
\beq
\frac{m_i^2 - m_j^2}{m_W^2}\,.
\eeq
So FCNCs probe directly the differences between the generations.

One can also see that $CP$ violation is related to irremovable phases of Yukawa
couplings.  Starting from a term in Eq.~(\ref{Lyuk}),
\beq
Y_{ij}\, \ov{\psi_{Li}}\, \phi\, \psi_{Rj}
  + Y_{ij}^*\, \ov{\psi_{Rj}}\, \phi^\dagger\, \psi_{Li}
~\stackrel{CP}{\longrightarrow}~
Y_{ij}\, \ov{\psi_{Rj}}\, \phi^\dagger\, \psi_{Li}
  + Y_{ij}^*\, \ov{\psi_{Li}}\, \phi\, \psi_{Rj}\,.
\eeq
The two expressions are identical if and only if a basis for the quark fields
can be chosen such that $Y_{ij} = Y_{ij}^*$, i.e., that $Y_{ij}$ are real.

\paragraph{Counting flavor parameters}

Most parameters of the SM (and also of many of its extensions) are related to
flavor.  In the CKM matrix, due to unitarity, 9 complex elements depend on 9
real parameters.  Of these 5 phases can be absorbed by redefining the quark
fields, leaving 4 physical parameters, 3 mixing angles and 1 $CP$ violating
phase.  This is the only source of $CP$ violation in flavor-changing transitions
in the SM.

A more general way to account for all flavor parameters is to consider that the
two Yukawa matrices, $Y_{i,j}^{u,d}$ in Eq.~(\ref{Lyuk}), contain 18 real and 18
imaginary parameters.  They break the global $[U(3)]^3 \to U(1)_B$, see
Eq.~(\ref{quarkglobal}), so there are 26 broken generators (9 real and 17
imaginary).  This leaves 10 physical quark flavor parameters: 9 real ones (the 6
quark masses and 3 mixing angles) and 1 complex $CP$ violating
phase.\footnote{Show that for $N$ generations, the CKM matrix depends on
$N(N-1)/2$ mixing angles and $(N-1)(N-2)/2$ $CP$ violating phases.  So the
2-generation SM conserves $CP$.}

\paragraph{Neutrino masses}

How does lepton flavor differ?  With the particle content in Eq.~(\ref{reps}),
it is not possible to write down a renormalizable mass term for neutrinos.  It
would require introducing a $\nu_R(1,1)_0$ field, a singlet under all SM gauge
groups, to be light, which is unexpected.  Such a particle is sometimes called a
sterile neutrino, as it has no SM interactions.  Whether there are such fields
can only be decided experimentally.

Viewing the SM as a low energy effective theory, there is a single type of
dimension-5 gauge invariant term made of SM fields,
\beq\label{numass}
{\cal L}_Y = 
  - \frac{Y^{ij}_\nu}{\Lambda_{\rm NP}}\, L_{Li}^I\, L_{Lj}^I\, \phi\, \phi\,.
\eeq
This term gives rise to neutrino masses and also violates lepton number.  Its
suppression cannot be the electroweak scale, $1/v$ (instead of $1/\Lambda_{\rm
NP}$), because such a term in the Lagrangian cannot be generated from SM fields
at arbitrary loop level, or even nonperturbatively.  [Eq.~(\ref{numass})
violates $B-L$, which is an accidental symmetry of the SM that is not
anomalous.]  The above mass term is called a Majorana mass, as it couples
$\ov{\nu_L}$ to $(\nu_L)^c$ instead of $\nu_R$ [the latter occurs for Dirac mass
terms, see Eq.~(\ref{mass})].  The key distinction is whether lepton number is
violated or conserved.  In the presence of Eq.~(\ref{numass}) and the charged
lepton Yukawa coupling in the last term in Eq.~(\ref{Lyuk}), the global $U(3)_L
\times U(3)_\ell$ symmetry is completely broken, and the counting of lepton
flavor parameters is\footnote{Show that the Yukawa matrix in Eq.~(\ref{numass})
is symmetric, $Y^{ij}_\nu = Y^{ji}_\nu$.  Derive that for $N$ such generations
there are $N(N-1)/2$ $CP$ violating phases.}
\beq
(12+18 \mbox{ couplings}) - (18 \mbox{ broken sym.})
  ~\Rightarrow~ 12 \mbox{ physical parameters\,.}
\eeq
These are the 6 masses, 3 mixing angles, and 3 $CP$ violating phases, of which
one is the analog of the CKM phase measurable in oscillation experiments, while
two additional ``Majorana phases" only contribute to lepton number violating
processes, such as neutrinoless double beta decay.\footnote{Can you think of
ways to get sensitivity to another linear combination of the two $CP$ violating
Majorana phases, besides the one that enters neutrinoless double beta decay?}

\paragraph{The CKM matrix}

Quark mixing is observed to be approximately flavor diagonal.  The Wolfenstein
parametrization conveniently exhibits this,
\begin{equation}\label{ckmdef}
V_{\rm CKM} = \left( \begin{matrix} V_{ud} & V_{us} & V_{ub} \\
  V_{cd} & V_{cs} & V_{cb} \\
  V_{td} & V_{ts} & V_{tb} \end{matrix} \right)
= \left( \begin{matrix} 1-\frac{1}{2}\lambda^2 & \lambda & A\lambda^3(\rho-i\eta) \\
  -\lambda & 1-\frac{1}{2}\lambda^2 & A\lambda^2 \\
  A\lambda^3(1-\rho-i\eta) & -A\lambda^2 & 1 \end{matrix} \right) + \ldots \,,
\end{equation}
where $\lambda \simeq 0.23$ may be viewed as an ``expansion parameter". It is a
useful book-keeping of the magnitudes of the CKM matrix elements, but it hides
which combination of CKM elements are phase-convention independent.  Sometimes
it can be useful to think of $V_{ub}$ and $V_{td}$ as the ones with ${\cal
O}(1)$ $CP$ violating phases, but it is important that any $CP$ violating
observable in the SM must depend on at least four CKM elements.\footnote{Prove
this statement.  Are there constraints on which four?}

\begin{figure}[t]
\centerline{\includegraphics[width=5.4cm]{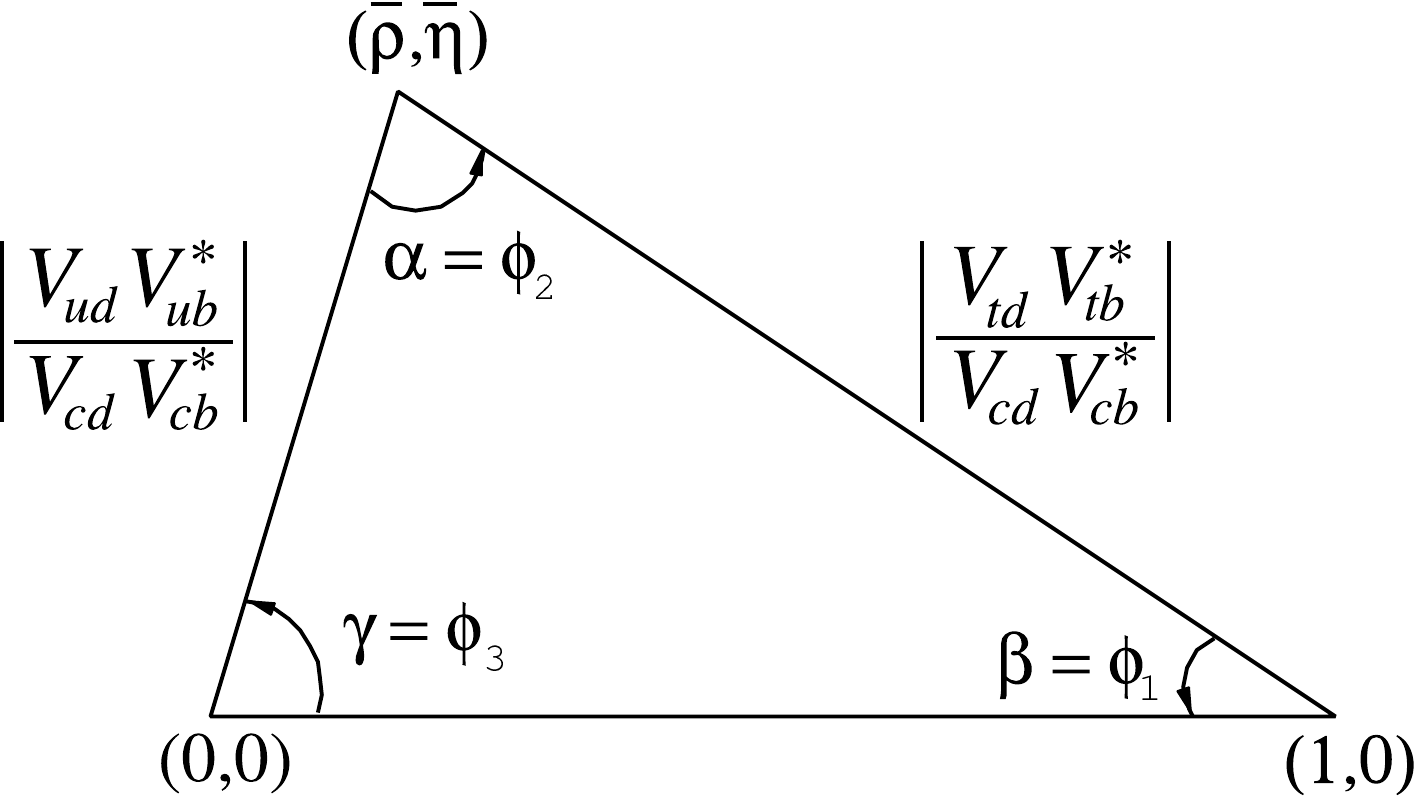}}
\caption{The unitarity triangle.}
\label{fig:UT}
\end{figure}

In any case, the interesting question is not primarily measuring CKM elements,
but testing how precisely the SM description of flavor and $CP$ violation
holds.  This can be done by ``redundant" measurements, which in the SM relate to
some combination of flavor parameters, but are sensitive to different BSM
physics, thus testing for (in)consistency.  Since there are many experimental
constraints, a simple way to compare different measurements can be very useful. 
Recall that CKM unitarity implies
\beq
\sum_k V_{ik} V_{jk}^* = \sum_k V_{ki} V_{kj}^* = \delta_{ij}\,,
\eeq
and the 6 vanishing relations can be represented as triangles in a complex
plane.  The most often used such ``unitarity triangle" (shown in
Fig.~\ref{fig:UT}) arises from the scalar product of the 1st and 3rd columns,
\beq
V_{ud}\, V_{ub}^* + V_{cd}\, V_{cb}^* + V_{td}\, V_{tb}^* = 0 \,.
\eeq
(Unitarity triangles constructed from neighboring columns or rows are
``squashed".)  We define the $\alpha$, $\beta$, $\gamma$ angles of this
triangle, and two more,
\beqa\label{angledef}
\alpha &\equiv& \arg\left(-{V_{td}V_{tb}^*\over V_{ud}V_{ub}^*}\right), \quad
  \beta \equiv \arg\left(-{V_{cd}V_{cb}^*\over V_{td}V_{tb}^*}\right), \quad
  \gamma \equiv \arg\left(-{V_{ud}V_{ub}^*\over V_{cd}V_{cb}^*}\right), \nn\\*
\beta_s &\equiv& \arg\left( -\frac{V_{ts}V_{tb}^*}{V_{cs}V_{cb}^*}\right), \quad
  \beta_K \equiv \arg\left( -\frac{V_{cs}V_{cd}^*}{V_{us}V_{ud}^*}\right).
\eeqa
On different continents the $\phi_1 = \beta$, $\phi_2 = \alpha$, $\phi_3
= \gamma$, and/or the $\phi_s = -2\beta_s$  notations are used.  Here $\beta_s$
($\beta_K$), of order $\lambda^2$ ($\lambda^4$), is the small angle of a
``squashed" unitarity triangle obtained by multiplying the 2nd column of the CKM
matrix with the 3rd (1st) column.

\begin{figure}[t]
\centerline{\includegraphics[width=.7\textwidth]{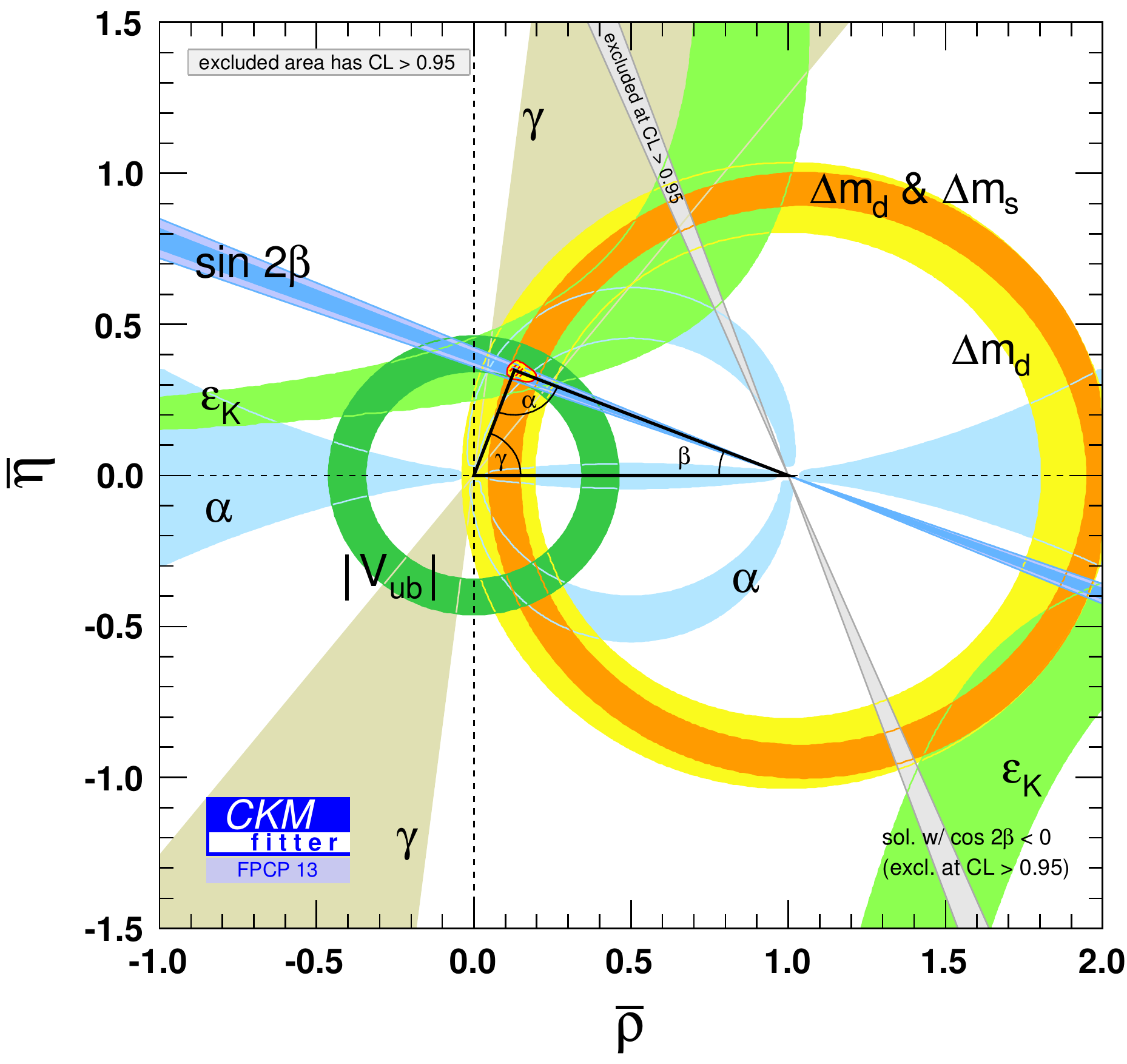}}
\caption{The SM CKM fit, and individual constraints (colored regions show
95\%~CL).}
\label{fig:SMCKMfit}
\end{figure}

The magnitudes of CKM elements determine the sides of the unitarity triangle.  They
are mainly extracted from semileptonic and leptonic $K$ and $B$ decays, and
$B_{d,s}$ mixing.  Any constraint which renders the area of the unitarity
triangle nonzero, such as angles, has to measure $CP$ violation.  Some of the
most important constraints are shown in Fig.~\ref{fig:SMCKMfit}, together with
the CKM fit in the SM.  (Using $\rhobar,\, \etabar$ instead of $\rho,\, \eta$
simply corresponds to a small modification of the parametrization, to keep
unitarity exact.)

\paragraph{The low energy effective field theory (EFT) viewpoint} 

\begin{figure}[t]
\centerline{\includegraphics[width=7.5cm]{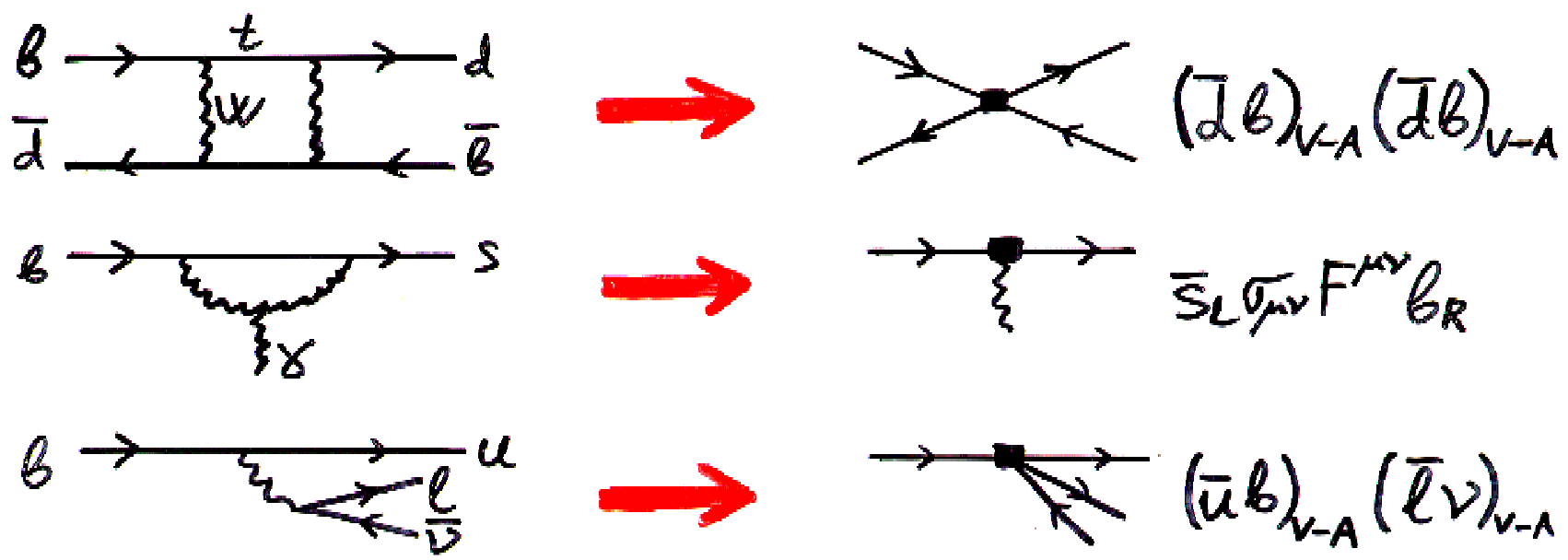}}
\caption{Diagrams at the electroweak scale (left) and operators at the scale
$m_b$ (right).}
\label{scannedfig}
\end{figure}

At the few GeV scale, relevant for $B$, $D$, and some $K$ decays, all flavor
changing processes (both tree and loop level) are mediated by dozens of higher
dimension local operators.  They arise from integrating out heavy particles, $W$
and $Z$ bosons and the $t$ quark in the SM, or not yet observed heavy states
(see Fig.~\ref{scannedfig}).  Since the coefficients of a large number of
operators depend on just a few parameters in the SM, there are many correlations
between decays of hadrons containing $s$, $c$, $b$ quarks, which NP may
violate.  From this point of view there is no difference between flavor-changing
neutral currents and $\Delta F=1$ processes, as all flavor-changing processes
are due to~heavy particles with masses $\gg m_{s,c,b}$.  Thus, one can test the
SM in many ways by asking (i) does NP modify the coefficients of dimension-6
operators? (ii) does NP generate operators absent in the SM (e.g., right-handed
couplings)?

\paragraph{Neutral meson mixing} 

Let us first sketch a back-of-an-envelope estimate of the mass difference in
$K^0$\,--\,$\K0bar$ mixing.  In the SM,
\beq
\Delta m_K \sim \alpha_w^2\, |V_{cs}V_{cd}|^2\,
{m_c^2-m_u^2\over m_W^4}\, f_K^2\, m_K\,.
\eeq
The result is suppressed by CKM angles, a loop factor, the weak coupling, and
the GIM mechanism.  If a heavy particle, $X$, contributes ${\cal O}(1)$
to $\Delta m_K$,
\beq
\bigg|{\Delta m_K^{(X)}\over \Delta m_K^{\rm (exp)}}\bigg|
\sim \bigg|{g^2\, \lqcd^3\over M_X^2\, \Delta m_K^{\rm (exp)}}\bigg|
\quad\Rightarrow\quad \frac{M_X}g \gtrsim\, 2\times 10^3\, \TeV\,.
\eeq
So even TeV-scale particles with loop-suppressed couplings $[g \sim {\cal
O}(10^{-3})]$ can give observable effects.  This illustrates that flavor physics
measurements indeed probe the TeV scale if NP has SM-like flavor structure, and
much higher scales if the NP flavor structure is generic.

A more careful evaluation of the bounds in all four neutral meson systems is
shown in Table~\ref{tab:mixingscales}.  (See Sec.~\ref{sec2} for the definitions
of the observables in the $B$ meson systems.)  If $\Lambda = {\cal O}(1\,\TeV)$
then $C\ll 1$, and if $C = {\cal O}(1)$ then $\Lambda \gg 1\,\TeV$.  The bounds
are weakest for $B_{(s)}$ mesons, as mixing is the least suppressed in the SM in
that case.  The bounds on many NP models are the strongest from $\Delta m_K$ and
$\epsilon_K$, since so are the SM suppressions.  These are built into NP models
since the 1970s, otherwise the models are immediately excluded.  In the SM,
larger FCNCs and $CP$ violating effects occur in $B$ mesons, which can be
measured precisely.  In many BSM models the 3rd generation is significantly
different than the first two, motivated by the large top Yukawa, and may give
larger signals in the $B$ sector.

\begin{table}[tb]\tabcolsep 4pt
\tbl{Bounds on some $\Delta F=2$ operators, $(C/\Lambda^2)\,
{\cal O}$, with ${\cal O}$ given in the first column. The bounds on $\Lambda$
assume $C=1$, the bounds on $C$ assume $\Lambda=1$\,TeV.
(From Ref.~\cite{Isidori:2010kg}.)}
{\begin{tabular}{@{}c|cc|cc|c@{}}
\hline\hline
\multirow{2}{*}{Operator} &
  \multicolumn{2}{c|}{Bound on $\Lambda$\,[TeV]~($C=1$)} &
  \multicolumn{2}{c|}{Bound on $C$~($\Lambda=1$\,TeV)} & 
  \multirow{2}{*}{Observables}\\
&   Re  & Im  &  Re  &  Im  &  \\
\hline
\raisebox{0pt}[9pt][0pt]{$(\bar s_L \gamma^\mu d_L )^2$}
  &  $9.8 \times 10^{2}$  &  $1.6 \times 10^{4}$
  &$9.0 \times 10^{-7}$& $3.4 \times 10^{-9}$ & $\Delta m_K$; $\epsilon_K$ \\
($\bar s_R\, d_L)(\bar s_L d_R$)   & $1.8 \times 10^{4}$& $3.2 \times 10^{5}$
  &$6.9 \times 10^{-9}$& $2.6 \times 10^{-11}$ &  $\Delta m_K$; $\epsilon_K$ \\
\hline
\raisebox{0pt}[9pt][0pt]{$(\bar c_L \gamma^\mu u_L )^2$}
  &  $1.2 \times 10^{3}$  &  $2.9 \times 10^{3}$
&  $5.6 \times 10^{-7}$& $1.0 \times 10^{-7}$ & $\Delta m_D$; $|q/p|, \phi_D$ \\
($\bar c_R\, u_L)(\bar c_L u_R$)   & $6.2 \times 10^{3}$& $1.5 \times 10^{4}$
&  $5.7 \times 10^{-8}$& $1.1 \times 10^{-8}$ &  $\Delta m_D$; $|q/p|, \phi_D$\\
\hline
\raisebox{0pt}[9pt][0pt]{$(\bar b_L \gamma^\mu d_L )^2$}
  &  $6.6 \times 10^{2}$ & $9.3 \times 10^{2}$ &  $2.3 \times 10^{-6}$ &
  $1.1 \times 10^{-6}$ & $\Delta m_{B_d}$; $S_{\psi K_S}$  \\
($\bar b_R\, d_L)(\bar b_L d_R)$  &   $2.5 \times 10^{3}$ & $3.6
\times 10^{3}$ &  $3.9 \times 10^{-7}$ &  $1.9 \times 10^{-7}$
&   $\Delta m_{B_d}$; $S_{\psi K_S}$ \\
\hline 
\raisebox{0pt}[9pt][0pt]{$(\bar b_L \gamma^\mu s_L )^2$}
  &  $1.4 \times 10^2$ & $2.5 \times 10^2$  &
  $5.0\times10^{-5}$  &  $1.7\times10^{-5}$  & $\Delta m_{B_s}$; $S_{\psi\phi}$ \\
($\bar b_R \,s_L)(\bar b_L s_R)$  &  $4.8 \times 10^2$ & $8.3 \times 10^2$ &
 $8.8\times10^{-6}$ &  $2.9\times10^{-6}$  & $\Delta m_{B_s}$; $S_{\psi\phi}$ \\ 
\hline\hline
\end{tabular}}
\label{tab:mixingscales}
\end{table}

\paragraph{A few more words on kaons}

With recent lattice QCD progress on $B_K$ and $f_K$~\cite{Aoki:2013ldr},
$\epsilon_K$ has become a fairly precise constraint on the SM.  However,
$\epsilon'_K$ is notoriously hard to calculate, involving cancellation between
two comparable terms, each with sizable uncertainties.  (Lattice QCD
calculations of the hadronic matrix elements for $\epsilon'_K$ may be reliably
computed in the future.)  At present, we cannot prove nor rule out that a large
part of the observed value of $\epsilon'_K$ is due to BSM.  Thus, to test $CP$
violation, one had to consider other systems; it was realized in the 1980s that
many precise measurements of $CP$ violation are possible in $B$ decays.

\begin{figure}[b]
\centerline{\includegraphics*[width=.5\textwidth]{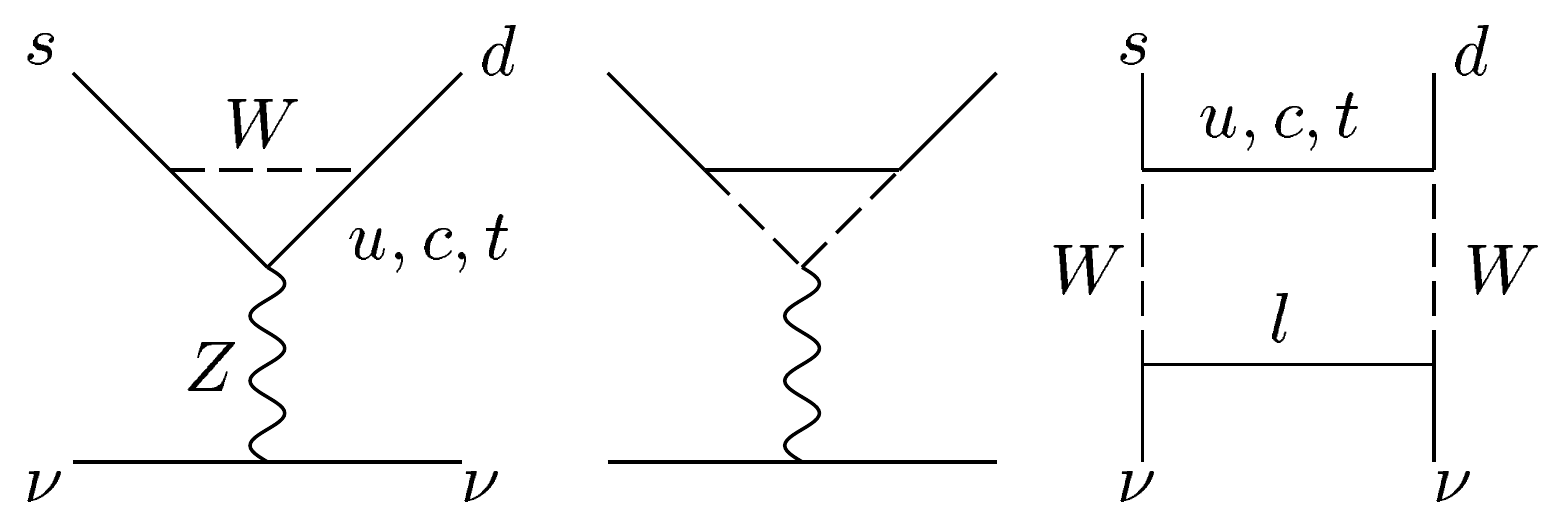}}
\vspace*{-4pt}
\caption{Diagrams contributing to $K\to \pi\nu\bar\nu$ decay.}
\label{fig:kpinunu}
\end{figure}

In the kaon sector, precise calculations of rare decays involving neutrinos (see
Fig.~\ref{fig:kpinunu}) are possible, and the SM predictions
are~\cite{Buras:2015qea}
\beq
{\cal B}(K^+ \!\to \pi^+\nu\bar\nu) = (8.4\pm1.0)\times 10^{-11}, \quad
{\cal B}(K^0_L \to \pi^0\nu\bar\nu) = (3.4\pm0.6)\times 10^{-11}.
\eeq 
The $K_L^0$ decay is $CP$ violating, and therefore it is under especially good
theoretical control, since it is determined by the top quark loop contribution,
and the $CP$ conserving charm quark contribution is absent (which enters $K^+
\to \pi^+\nu\bar\nu$, and is subject to some hadronic uncertainty).

The E787/E949 measurement is ${\cal B}(K\to\pi^+ \nu\bar\nu) =
(17.3^{+11.5}_{-10.5})\times 10^{-11}$~\cite{Artamonov:2008qb}, whereas in the
$K_L$ mode the experimental upper bound is still many times the SM rate.  NA62
at CERN aims to measure the $K^+$ rate with 10\% uncertainty, and will start to
have dozens of events in 2015.  The $K_L$ mode will probably be first observed
by the KOTO experiment at J-PARC.

\section{Theory of Some Important $B$ Decays}
\addcontentsline{toc}{}{Theory of Some Important $B$ Decays}
\label{sec2}

Studying FCNC and $CP$ violation is particularly interesting in $B$ meson
decays, because many measurements are possible with clean interpretations.

The main theoretical reasons are: (i) $t$ quark loops are neither GIM nor CKM
suppressed; (ii) large $CP$ violating effects are possible; (iii) some of the
hadronic physics is understandable model independently ($m_b \gg \lqcd$).

The main experimental reasons are: (i) the long $B$ lifetime (small $|V_{cb}|$);
(ii) the $\Upsilon(4S)$ is a clean source of $B$ mesons at $e^+e^-$ colliders;
(iii) for $B_d$, the ratio $\Delta m/\Gamma = {\cal O}(1)$.

\paragraph{Neutral meson mixing formalism}

\begin{figure}[t]
\centerline{\includegraphics*[height=.15\textwidth]{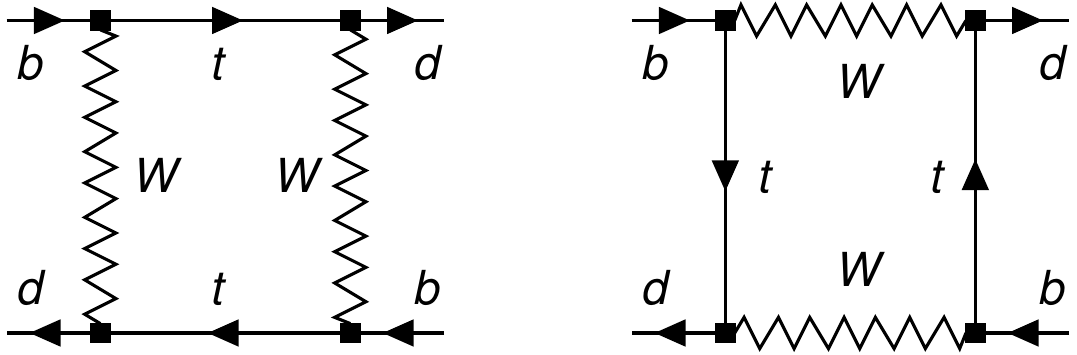}
\hspace{.75cm} \raisebox{.7cm}{$\Longrightarrow$} \hspace{.5cm}
\includegraphics*[height=.15\textwidth]{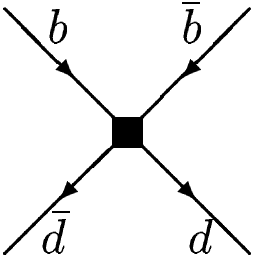}}
\caption{Left: box diagrams that give rise to the $B^0-\B0bar$ mass difference;
Right: operator in the effective theory below $m_W$ whose $B$ meson matrix
element determines $\Delta m_B$.}
\label{fig:boxes}
\end{figure}

Similar to neutral kaons, there are two neutral $B^0$ meson flavor eigenstates,
\beq
|B^0\rangle = |\bar b\, d\rangle \,, \qquad
  |\B0bar\rangle = |b\, \bar d\rangle \,.
\eeq
They mix in the SM due to weak interactions (see Fig.~\ref{fig:boxes}).  The
time evolutions of the two states are described by the Schr\"odinger equation,
\beq
i\, {\d \over \d t} \left( \begin{matrix}|B^0(t)\rangle\\
  |\B0bar(t)\rangle \end{matrix}\right) 
= \Big(M - {i\over2}\,\Gamma\Big)\! 
  \left( \begin{matrix}|B^0(t)\rangle \\ 
  |\B0bar(t)\rangle \end{matrix}\right) \,,
\eeq
where the mass ($M$) and the decay ($\Gamma$) mixing matrices are $2\times2$
Hermitian matrices.   $CPT$ invariance implies $M_{11} = M_{22}$ and
$\Gamma_{11} = \Gamma_{22}$.   The heavier and lighter mass eigenstates are the
eigenvectors of $M - i\Gamma/2$,
\beq\label{defpq}
|B_{H,L}\rangle = p\, |B^0\rangle \mp q\, |\B0bar\rangle\,,
\eeq
and their time dependence is 
\beq\label{timedep}
|B_{H,L}(t)\rangle = e^{-(im_{H,L} + \Gamma_{H,L}/2)t}\, |B_{H,L}\rangle\,.
\eeq
Here $\Delta m \equiv m_H - m_L$ and $\Delta\Gamma = \Gamma_L - \Gamma_H$ are
the mass and width differences.  This defines $\Delta m$ to be positive, but the
sign of $\Delta\Gamma$ is physical.  Note that $m_{H,L}$ ($\Gamma_{H,L}$) are
not the eigenvalues of $M$ ($\Gamma$).\footnote{Derive that the time evolutions
of mesons that are $B^0$ and $\B0bar$ at $t=0$ are given by
\beq\label{timev}
|B^0(t)\rangle = g_+ (t)\, |B^0\rangle
  + \frac{q}{p}\, g_- (t)\, |\B0bar\rangle\,, \qquad
|\B0bar(t)\rangle = \frac{p}{q}\, g_- (t)\, |B^0\rangle
  + g_+(t)\, |\B0bar\rangle\,,
\eeq
where, denoting $m=(m_H+m_L)/2$ and $\Gamma=(\Gamma_H+\Gamma_L)/2$,
\beqa\label{qpm}
g_+ (t) &=& e^{-i t (m - i\Gamma/2)}\,
  \bigg( \cosh\frac{\Delta\Gamma\, t}{4}\, \cos\frac{\Delta m\, t}{2} -   
  i \sinh\frac{\Delta\Gamma\, t}{4}\, \sin \frac{\Delta m \, t}{2} \bigg)\,,\nn\\
g_- (t) &=& e^{-i t (m - i\Gamma/2)}\,
  \bigg(- \sinh\frac{\Delta\Gamma\, t}{4}\, \cos \frac{\Delta m \, t}{2} +
  i \cosh\frac{\Delta\Gamma\, t}4\, \sin \frac{\Delta m \, t}{2} \bigg)\,.
\eeqa}
The off-diagonal elements, $M_{12}$ and $\Gamma_{12}$, arise from virtual and
on-shell intermediate states, respectively.  In the SM, $M_{12}$ is dominated by
the top-quark box diagrams in Fig.~\ref{fig:boxes}.  Thus, $M_{12}$ is
determined by short-distance physics, it is calculable with good accuracy, and
is sensitive to high scales.  (This is the complication for $D$ mixing: the
$W$ can always be shrunk to a point, but the $d$ and $s$ quarks in the box
diagrams cannot, so long-distance effects are important.)  The width difference
$\Gamma_{12}$ is determined by on-shell states to~which both $B^0$ and
$\B0bar$ can decay, corresponding to $c$ and $u$ quarks in the box diagrams.

The solution of the eigenvalue equation is
\beqa\label{eigenstates}
(\Delta m)^2 - \frac{(\Delta\Gamma)^2}{4} 
  &=& 4\, |M_{12}|^2 - |\Gamma_{12}|^2\,, \qquad
\Delta m\, \Delta\Gamma = -4\, {\rm Re} (M_{12} \Gamma_{12}^*)\,, \nn\\
{q\over p} &=& - \frac{\Delta m + i\, \Delta\Gamma/2}{2M_{12} -i\, \Gamma_{12}}
  = - \frac{2 M_{12}^* -i\, \Gamma_{12}^*}{\Delta m + i\, \Delta\Gamma/2}\,.
\eeqa
The physical observables that are measurable in neutral meson mixing are
\beq
x = \frac{\Delta m}{\Gamma}\,, \qquad
  y = \frac{\Delta\Gamma}{2\Gamma}\,,\qquad
  \bigg|\frac{q}{p}\bigg|-1\,.
\eeq
The orders of magnitudes of the SM predictions are shown in
Table~\ref{mixingtable}.  That $x\neq 0$ is established in the $K$, $B$, and
$B_s$ mixing; $y\neq 0$ in the $K$, $D$, and $B_s$ mixing; $|q/p|\neq1$ in
$K$ mixing.  The significance of $x_D\neq0$ is $\sim2\sigma$, and in $B_{d,s}$
mixing there is an unconfirmed D\O\ signal for $|q/p|\neq1$; more~below.

\begin{table}[tb]
\tabcolsep 12pt
\tbl{Orders of magnitudes of the SM predictions for mixing
parameters.  The uncertainty of $(|q/p|-1)_D$ is especially large.}
{\begin{tabular}{c|ccc}
\hline\hline
\raisebox{0pt}[9pt][0pt]{meson}  &  $x = \Delta m/\Gamma$  & 
  $y=\Delta\Gamma/(2\Gamma)$  &  $|q/p|-1$ \\
\hline
\raisebox{0pt}[9pt][0pt]{$K$}  &  $1$  &  $1$  &  $10^{-3}$ \\
$D$  &  $10^{-2}$  &  $10^{-2}$  &  $10^{-3}$ \\
$B_d$  &  $1$  &  $10^{-2}$  &  $10^{-4}$ \\
$B_s$  &  $10^1$  &  $10^{-1}$  &  $10^{-5}$ \\
\hline\hline
\end{tabular}}
\label{mixingtable}
\end{table}

Simpler approximate solutions can be obtained expanding about the limit
$|\Gamma_{12}| \ll |M_{12}|$.  This is a good approximation in both $B_d$ and
$B_s$ systems.  $|\Gamma_{12}| < \Gamma$ always holds, because $\Gamma_{12}$
arises from decays to final states common to $B^0$ and $\B0bar$.  For $B_s$
mixing the world average is $\Delta \Gamma_s / \Gamma_s = 0.138 \pm
0.012$~\cite{hfag:2014}, while $\Delta \Gamma_d$ is expected to be $\sim\!20$
times smaller and is not yet measured.  Up to higher order terms in
$|\Gamma_{12} / M_{12}|$, Eqs.~(\ref{eigenstates}) become
\beqa\label{limes}
\Delta m &=& 2\, |M_{12}|\,, \qquad
  \Delta\Gamma = -2\, {{\rm Re} (M_{12} \Gamma_{12}^*)\over |M_{12}|}\,, \nn\\
{q\over p} &=& - {M_{12}^*\over |M_{12}|} \left(1 - \frac12\, {\rm Im}\,
  {\Gamma_{12}\over M_{12}} \right) ,
\eeqa
where we kept the second term in $q/p$, as it will be needed later.

\paragraph{$CP$ violation in decay}

This is any form of $CP$ violation that cannot be absorbed in a neutral meson
mixing amplitude (also called direct $CP$ violation).  It can occur in any
hadron decay, as opposed to those specific to neutral mesons discussed below. 
For a given final state, $f$, the $B\to f$ and $\Bbar\to \ov f$ decay amplitudes
can, in general, receive several contributions
\beq
A_f = \langle f | {\cal H} |B\rangle = 
  \sum_k A_k\, e^{i\delta_k}\, e^{i\phi_k} , \qquad
\ov{A}_{\ov f} = \langle \ov f | {\cal H} |\Bbar\rangle = 
  \sum_k A_k\, e^{i\delta_k}\, e^{-i\phi_k} .
\eeq
There are two types of complex phases.  Complex parameters in the Lagrangian
which enter a decay amplitude also enter the $CP$ conjugate amplitude but in
complex conjugate form.  In the SM such ``weak phases", $\phi_k$, only occur in
the CKM matrix.  Another type of phase is due to absorptive parts of decay
amplitudes, and gives rise to $CP$ conserving ``strong phases", $\delta_k$. 
These phases  arise from on-shell intermediate states rescattering into the
desired final state, and they are the same for an amplitude and its $CP$
conjugate.  The individual phases $\delta_k$ and $\phi_k$ are convention
dependent, but the phase differences, $\delta_i - \delta_j$ and $\phi_i -
\phi_j$, and therefore $|\ov{A}_{\ov{f}}|$ and $|A_f|$, are physical. Clearly,
if $|\ov{A}_{\ov{f}}| \neq |A_f|$ then $CP$ is violated; this is called $CP$
violation in decay, or direct $CP$ violation.\footnote{Derive that direct $CP$
violation requires interference of at least two contributing amplitudes with
different strong and weak phases, $|\ov A|^2 - |A|^2 = 4 A_1 A_2
\sin(\delta_1-\delta_2) \sin(\phi_1-\phi_2)$.}

There are many measurements of direct $CP$ violation.  While some give
strong constraints on NP models which evade the SM suppressions (e.g.,
$\epsilon'_K$, the first direct $CP$ violation measured with high significance),
at present no single direct $CP$ violation measurement gives a precise test of
the SM, due to the lack of reliable calculations of relevant strong phases.  For
all observations of direct $CP$ violation in a single decay mode, viewed in
isolation [see the caveat near Eq.~(\ref{directCPVtricky})], it is possible
that, say, half of the measured value is from BSM.  For $\epsilon'_K$, lattice
QCD may yield progress in the future.  In certain $B$ decays we may better
understand the implications of the heavy quark limit; so far $A_{K^+\pi^0} -
A_{K^+\pi^-} = 0.12 \pm 0.02$~\cite{hfag:2014}, the ``$K\pi$ puzzle", is poorly
understood.

\paragraph{$CP$ violation in mixing}

\begin{figure}[t]
\centerline{\includegraphics*[width=.6\textwidth]{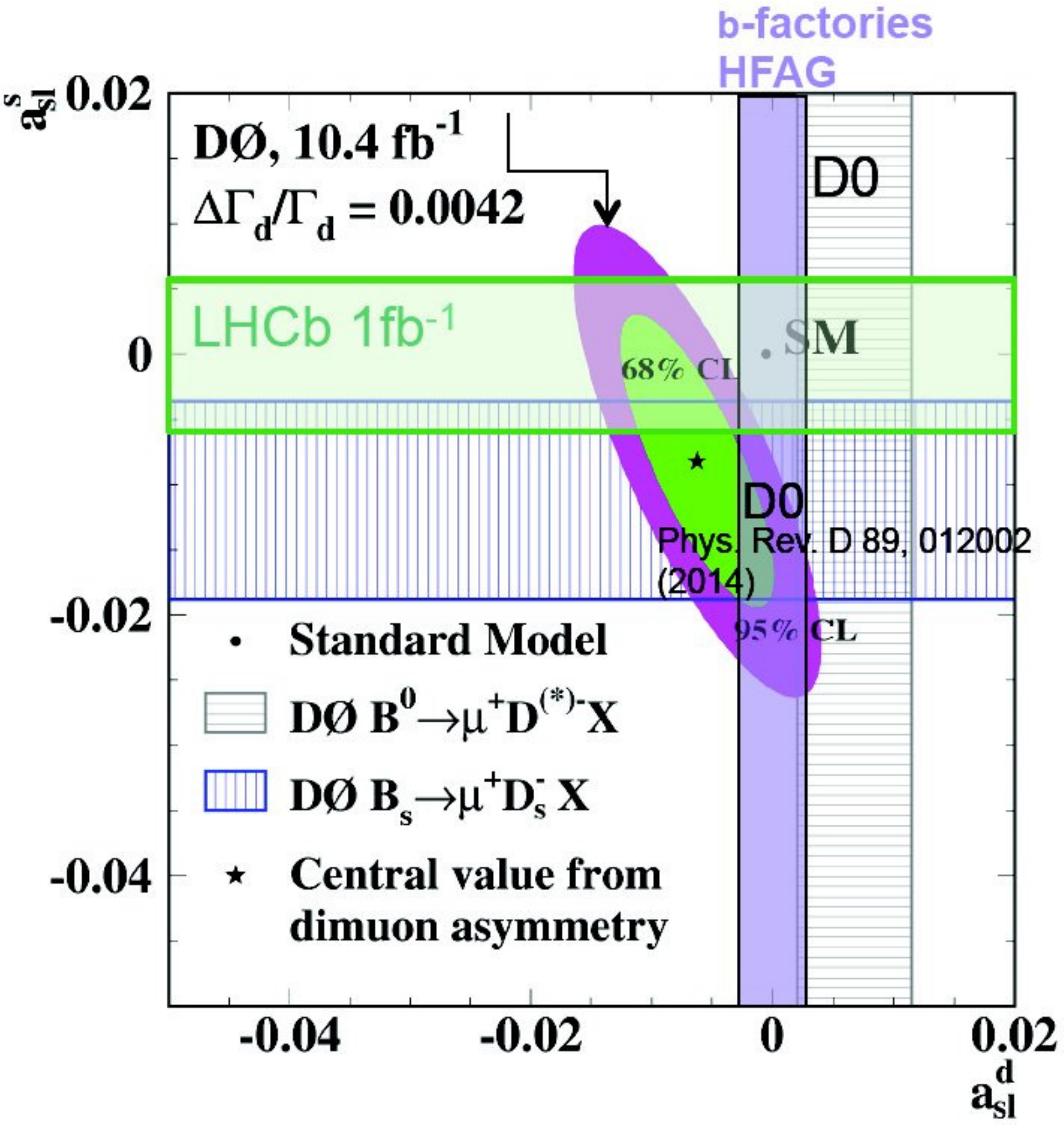}}
\caption{Status of $A_{\rm SL}$ measurements (from M.\ Artuso, talk at FPCP
2014).  The D\O\ result is in a $3.6\sigma$ tension with the SM expectation.}
\label{fig:ASL}
\end{figure}

If $CP$ were conserved, the mass and $CP$ eigenstates would coincide, and the
mass eigenstates would be proportional to $|B^0\rangle \pm |\B0bar\rangle$, up
to phases; i.e., $|q/p| = 1$ and ${\rm arg}(M_{12}/\Gamma_{12}) = 0$.  If $|q/p|
\neq 1$, then $CP$ is violated.  This is called $CP$ violation in mixing.  It
follows from Eq.~(\ref{defpq}) that $\langle B_H | B_L\rangle = |p|^2 - |q|^2$,
so if $CP$ is violated in mixing, the physical states are not orthogonal.  (This
illustrates again that $CP$ violation is a quantum mechanical effect, impossible
in a classical system.)  The simplest example is the $CP$ asymmetry in
semileptonic decay of neutral mesons to ``wrong sign" leptons
(Fig.~\ref{fig:ASL} summarizes the data),
\beq
A_{\rm SL}(t) = {\Gamma(\B0bar(t) \to \ell^+ X) - \Gamma(B^0(t) \to \ell^- X)
  \over \Gamma(\B0bar(t) \to \ell^+ X) + \Gamma(B^0(t) \to \ell^- X) }
= {1 - |q/p|^4 \over 1 + |q/p|^4} 
  \simeq {\rm Im}\, {\Gamma_{12}\over M_{12}} \,.
\eeq
To obtain the right-hand side, use Eqs.~(\ref{timev}) and (\ref{qpm})
for the time evolution, and Eq.~(\ref{limes}) for $|q/p|$.  In kaon decays this
asymmetry is measured~\cite{Angelopoulos:1998dv}, in agreement with the
SM prediction, $4\,{\rm Re}\, \epsilon_K$.  In $B_d$ and $B_s$ decays the
asymmetry is expected to be~\cite{Lenz:2014jya}
\beq\label{SMasl}
A_{\rm SL}^d \approx -4 \times 10^{-4} \,, \qquad
A_{\rm SL}^s \approx 2 \times 10^{-5} \,.
\eeq
The calculation of ${\rm Im} (\Gamma_{12} / M_{12})$ requires calculating
inclusive nonleptonic decay rates, which can be addressed using an operator
product expansion in the  $m_b \gg \lqcd$ limit.  Such a calculation has sizable
hadronic uncertainties, the details of which would lead to a long discussion. 
The constraints on new physics are significant
nevertheless~\cite{Laplace:2002ik}, as the $m_c^2/m_b^2$ suppression of $A_{\rm
SL}$ in the SM can be avoided in the presence of new physics.

\paragraph{$CP$ violation in the interference of decay with and without mixing}

A third type of $CP$ violation is possible when both $B^0$ and $\B0bar$ can
decay to a final state, $f$.  In the simplest cases, when $f$ is a $CP$
eigenstate, define
\beq
\lambda_f = \frac qp\, \frac{\ov A _f}{A_f} \,.
\eeq
If there is no direct $CP$ violation in a given mode, then $\ov A_f = \eta_f\,
\ov A_{\ov f}$, where $\eta_f = \pm 1$ is the $CP$ eigenvalue of $f$ [$+1$
($-1$) for $CP$-even (-odd) states].  This is useful, because $A_f$ and $\ov
A_{\ov f}$ are related by a $CP$ transformation.  If $CP$ were conserved, then
not only $|q/p| = 1$ and $|\ov A_{\ov f}/A_f| = 1$, but the relative phase
between $q/p$ and $\ov{A}_f/A_f$ also vanishes, hence $\lambda_f = \pm 1$.

\begin{figure}[t]
\centerline{\includegraphics*[width=.65\textwidth]{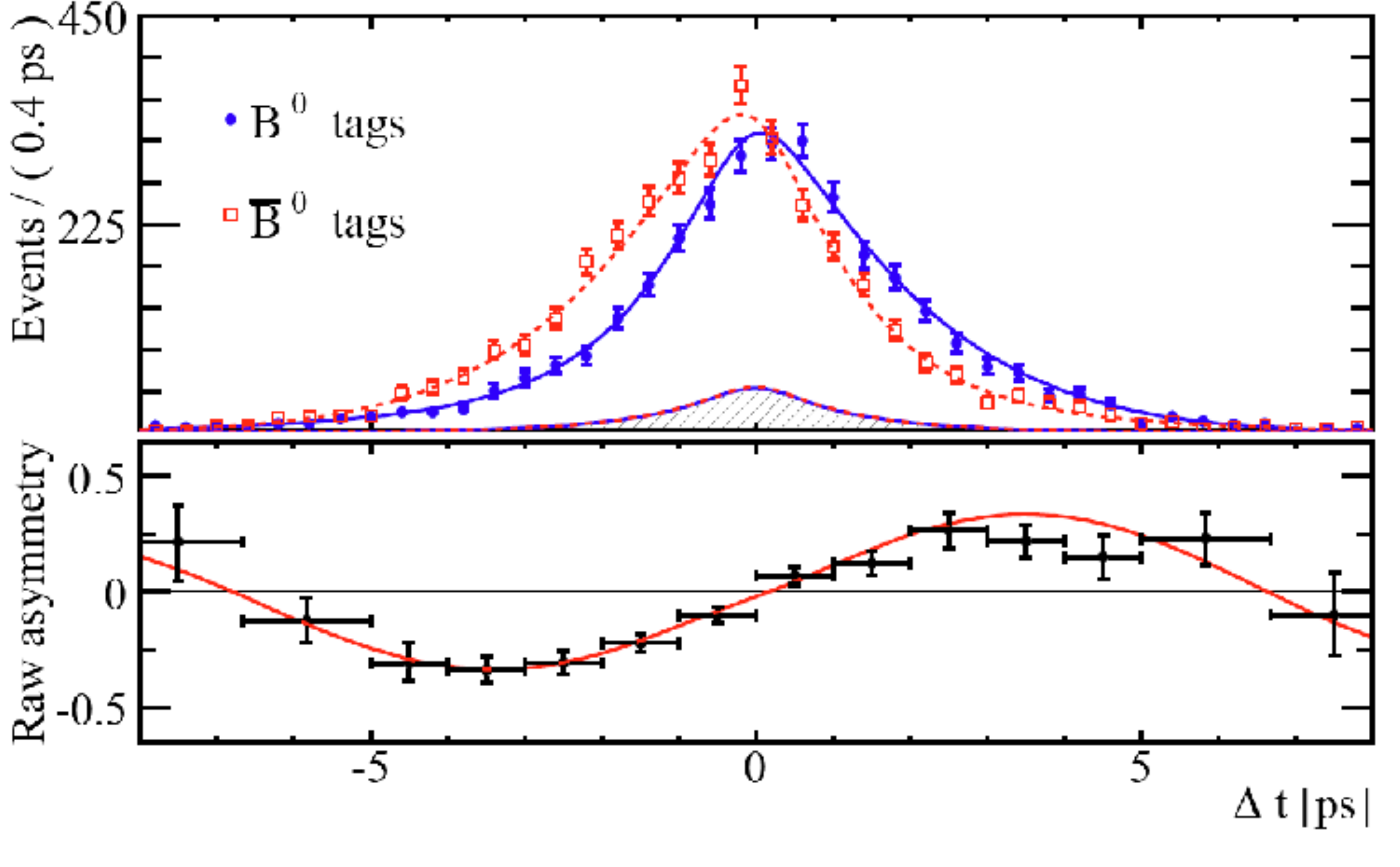}}
\vspace*{-4pt}
\caption{Time dependence of tagged $B\to \psi K$ decays (top);
$CP$ asymmetry (below)~\cite{Aubert:2009aw}.}
\label{fig:BpsiKSplot}
\end{figure}

The experimentally measurable $CP$ violating observable is\footnote{Derive the
$CP$ asymmetry in Eq.~(\ref{SCdef}) using Eq.~(\ref{timev})).  For extra credit,
keep $\Delta\Gamma \neq 0$.}
\beqa\label{SCdef}
a_f &=& { \Gamma[\B0bar(t)\to f] - \Gamma[B^0(t)\to f]\over
  \Gamma[\B0bar(t)\to f] + \Gamma[B^0(t)\to f] } \nn\\
&=& - {(1-|\lambda_f|^2) \cos(\Delta m\, t) 
  - 2\,{\rm Im}\,\lambda_f \sin(\Delta m\, t) \over 1+|\lambda_f|^2} \nn\\[2pt]
&\equiv& S_f \sin(\Delta m\, t) - C_f \cos(\Delta m\, t) \,,
\eeqa
where we have neglected $\Delta\Gamma$ (it is important in the $B_s$ system). 
The last line defines the $S$ and $C$ coefficients, which are fit to the
experimental data (see Fig.~\ref{fig:BpsiKSplot}).  If ${\rm Im}\lambda_f \neq
0$, then $CP$ violation arises in the interference between the decay $B^0 \to
f$, and mixing followed by decay, $B^0 \to \B0bar \to f$.

This asymmetry can be nonzero if any type of $CP$ violation occurs.  In
particular, in both the $B_d$ and $B_s$ systems $\big||q/p| - 1\big| < {\cal
O}(10^{-2})$ model independently, and it is much smaller in the SM [see,
Eq.~(\ref{SMasl})].  If, in addition, amplitudes with a single weak phase
dominate a decay, then $|\ov{A}_f/A_f| \simeq 1$, and ${\rm arg}\,
(\ov{A}_f/A_f)$ is just (twice) the weak phase, determined by short-distance
physics.  It is then possible that ${\rm Im} \lambda_f \neq 0$, $|\lambda_f|
\simeq 1$, and although we cannot compute the decay amplitude, we can extract
the weak phase difference between $B^0 \to f$ and $B^0 \to \B0bar \to f$ in a
theoretically clean way from the measurement of
\beq
a_f = {\rm Im} \lambda_f \sin(\Delta m\, t)\,.
\eeq

There is an interesting subtlety.  Consider two final states, $f_{1,2}$.  It is
possible that direct $CP$ violation in each channel, $|\lambda_{f_1}| - 1$ and
$|\lambda_{f_2}| - 1$, is unmeasurably small, but direct $CP$ violation is
detectable nevertheless.~If
\beq\label{directCPVtricky}
\eta_{f_1} {\rm Im}(\lambda_{f_1}) \neq \eta_{f_2} {\rm Im}(\lambda_{f_2}) \,,
\eeq
then $CP$ violation must occur outside the mixing amplitude, even though it may
be invisible in the data on any one final state.

\paragraph{$\sin2\beta$ from $B\to \psi K_{S,L}$}

This is one of the cleanest examples of $CP$ violation in the interference
between decay with and without mixing, and one of the theoretically cleanest
measurements of a CKM parameter.

\begin{figure}[tb]
\centerline{\includegraphics*[width=.3\textwidth]{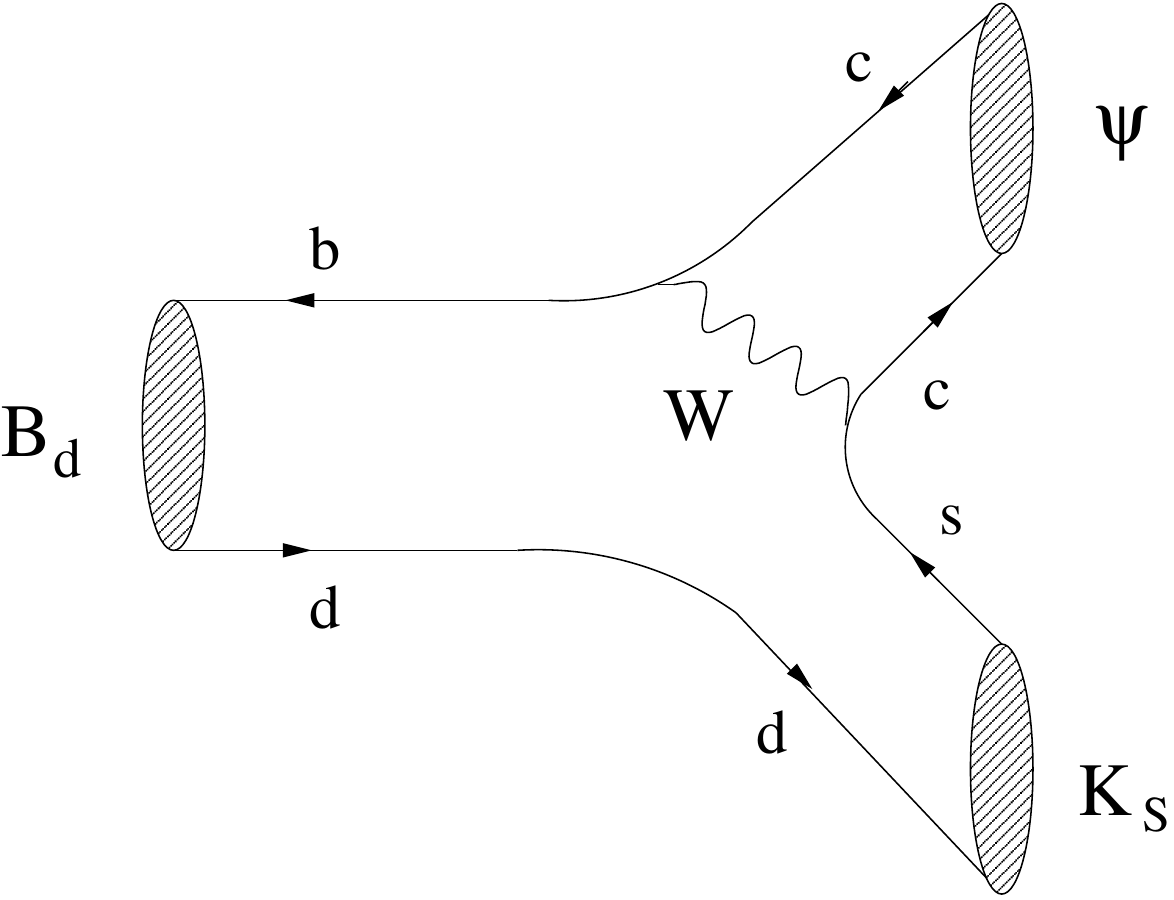}\hspace{1cm}
\includegraphics*[width=.3\textwidth]{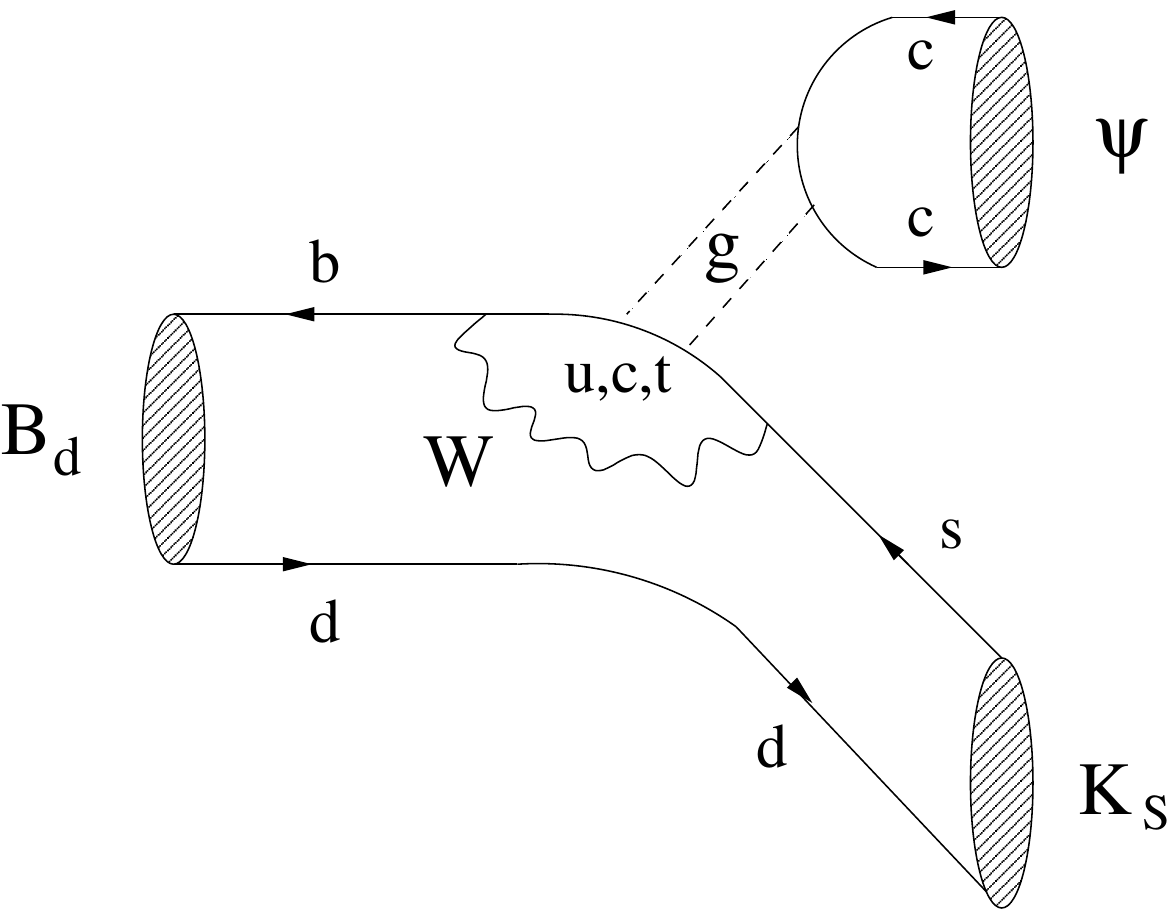}}
\caption{``Tree" (left) and ``penguin" (right) contributions to $B\to \psi
K_S$ (from Ref.~\cite{Fleischer:2002ys}).}
\label{fig:BpsiK}
\end{figure}

There are ``tree" and ``penguin" contributions to $B\to \psi K_{S,L}$, with
different weak and strong phases (see Fig.~\ref{fig:BpsiK}).  The tree
contribution is dominated by the $b\to c\bar c s$ transition, while there are
penguin contributions with three different combinations of CKM elements,
\beq
\ov A_T = V_{cb} V_{cs}^*\, T_{c\bar cs}\,, \qquad
  \ov A_P = V_{tb} V_{ts}^*\, P_t + V_{cb} V_{cs}^*\, P_c 
  + V_{ub} V_{us}^*\, P_u\,.
\eeq
($P_u$ can be defined to absorb the $V_{ub} V_{us}^*\, T_{u\bar us}$ ``tree"
contribution.)  We can rewrite the decay amplitude using $V_{tb} V_{ts}^* +
V_{cb} V_{cs}^* + V_{ub} V_{us}^* =0$ to obtain
\beqa\label{BpsiKamp}
\ov A &=& V_{cb} V_{cs}^*\, (T_{c\bar cs} + P_c - P_t)
  + V_{ub} V_{us}^*\, (P_u - P_t) \nn\\
&\equiv& V_{cb} V_{cs}^*\, T + V_{ub} V_{us}^*\, P\,,
\eeqa
where the second line defines $T$ and $P$.  Since $|(V_{ub} V_{us}^*) / (V_{cb}
V_{cs}^*)| \approx 0.02$, the $T$ amplitude with $V_{cb} V_{cs}^*$ weak phase
dominates.  Thus,
\beq\label{BpsiKlam}
\lambda_{\psi K_{S,L}} 
= \mp \bigg( { V_{tb}^* V_{td} \over V_{tb} V_{td}^*} \bigg)
  \bigg( {V_{cb} V_{cs}^* \over V_{cb}^* V_{cs}} \bigg)
  \bigg( {V_{cs} V_{cd}^* \over V_{cs}^* V_{cd}} \bigg) 
= \mp e^{-2i\beta} \,,
\eeq
and so ${\rm Im} \lambda_{\psi K_{S,L}} = \pm \sin2\beta$.  The first term is
the SM value of $q/p$ in $B_d$ mixing, the second is $\ov A/A$, the last one is
$p/q$ in the $K^0$ system, and $\eta_{\psi K_{S,L}} = \mp 1$.  Note that without
$K^0 - \K0bar$ mixing there would be no interference between $\B0bar\to \psi
\K0bar$ and $B^0\to \psi K^0$.  The accuracy of the relation between
$\lambda_{\psi K_{S,L}}$ and $\sin2\beta$ depends on  model dependent estimates
of $|P/T|$, which are below unity, so one expects it to be of order
\beq\label{BpsiKSCP}
\bigg| \frac{V_{ub} V_{us}^*}{V_{cb} V_{cs}^*}\, \frac{P}{T} \bigg|
  \lsim 10^{-2}\,.
\eeq
The absence of detectable direct $CP$ violation does not in itself bound this.
To fully utilize future LHCb and Belle~II data, better estimates are needed.

The first evidence for $CP$ violation outside the kaon sector was the BaBar and
Belle measurements of $S_{\psi K}$.  The current world average
is~\cite{hfag:2014}
\beq\label{s2b}
\sin2\beta = 0.682 \pm 0.019\,.
\eeq
This is consistent with other constraints, and shows that $CP$ violation in
quark mixing is an ${\cal O}(1)$ effect, which is simply suppressed in $K$
decays by small flavor violation suppressing the third generation's
contributions.

\paragraph{$\phi_s \equiv -2\beta_s$ from $B_s\to \psi\phi$}

\begin{figure}[t]
\centerline{\includegraphics*[width=.68\textwidth]{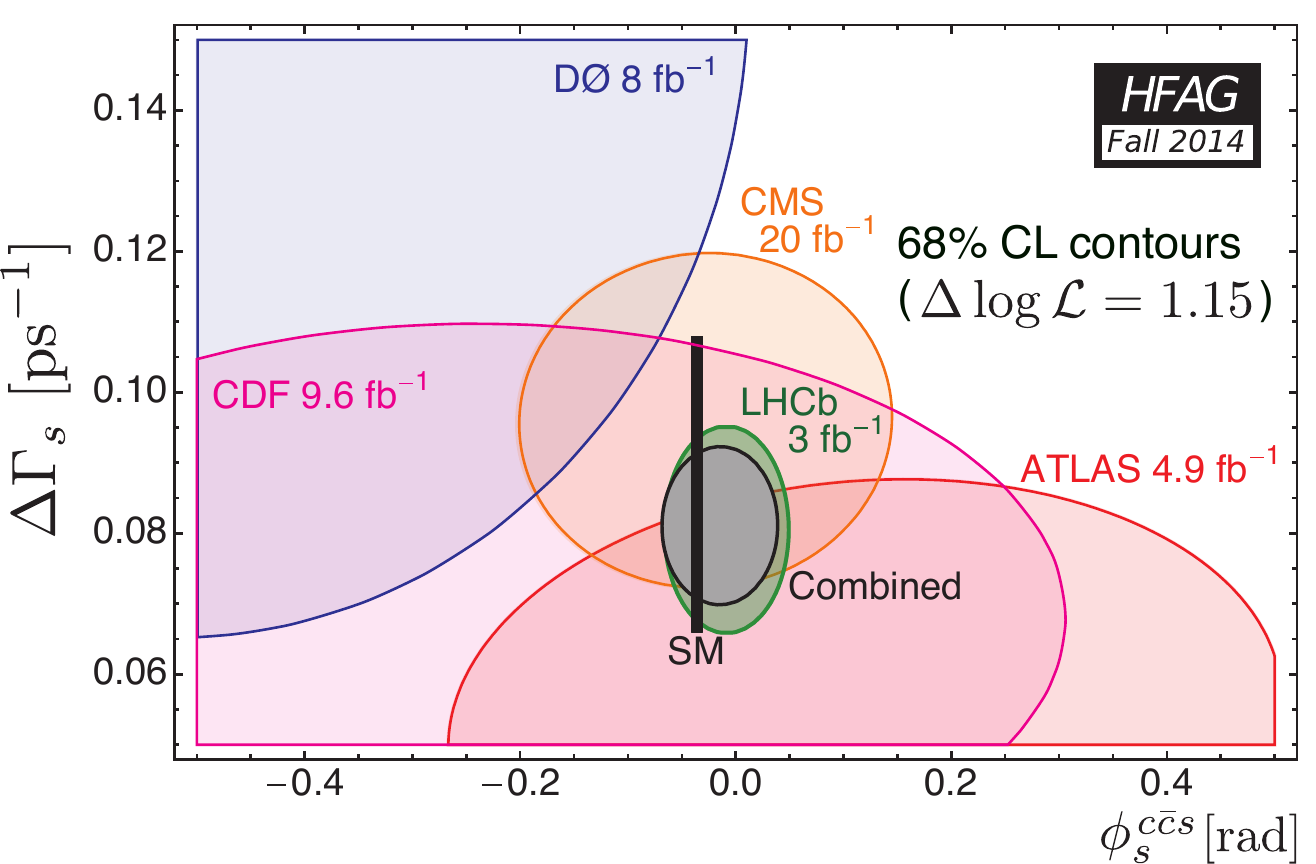}}
\vspace*{-4pt}
\caption{Measurements of $CP$ violation in $B_s\to \psi\phi$ and
$\Delta\Gamma_s$ (from Ref.~\cite{hfag:2014}).}
\label{fig:BsCPV}
\end{figure}

The analogous $CP$ asymmetry in $B_s$ decay, sensitive to BSM contributions to
$B_s$\,--\,$\Bbar_s$ mixing, is $B_s\to \psi\phi$.   Since the final state
consists of two vector mesons, it is a combination of $CP$-even ($L=0,2$) and
$CP$-odd ($L=1$) partial waves.  What is actually measured is the time-dependent
$CP$ asymmetry for each $CP$ component of the $\psi K^+K^-$ and $\psi
\pi^+\pi^-$ final states.   The SM prediction is suppressed compared to $\beta$
by $\lambda^2$, and is rather precise, $\beta_s =
0.0182^{+0.0007}_{-0.0006}$~\cite{Hocker:2001xe}. The latest LHCb result using
3\,fb$^{-1}$ data is~\cite{Aaij:2014zsa} (Fig.~\ref{fig:BsCPV} shows all
measurements)
\beq\label{s2bs}
\phi_s \equiv -2\beta_s = -0.010 \pm 0.039\,,
\eeq
which has an uncertainty approaching that of $2\beta$, suggesting that the
``room for new physics" in $B_s$ mixing is no longer larger than in $B_d$ (more
below).

\paragraph{``Penguin-dominated" measurements of \boldmath $\beta_{(s)}$}

\begin{figure}[t]
\centerline{\includegraphics*[height=.2\textwidth]{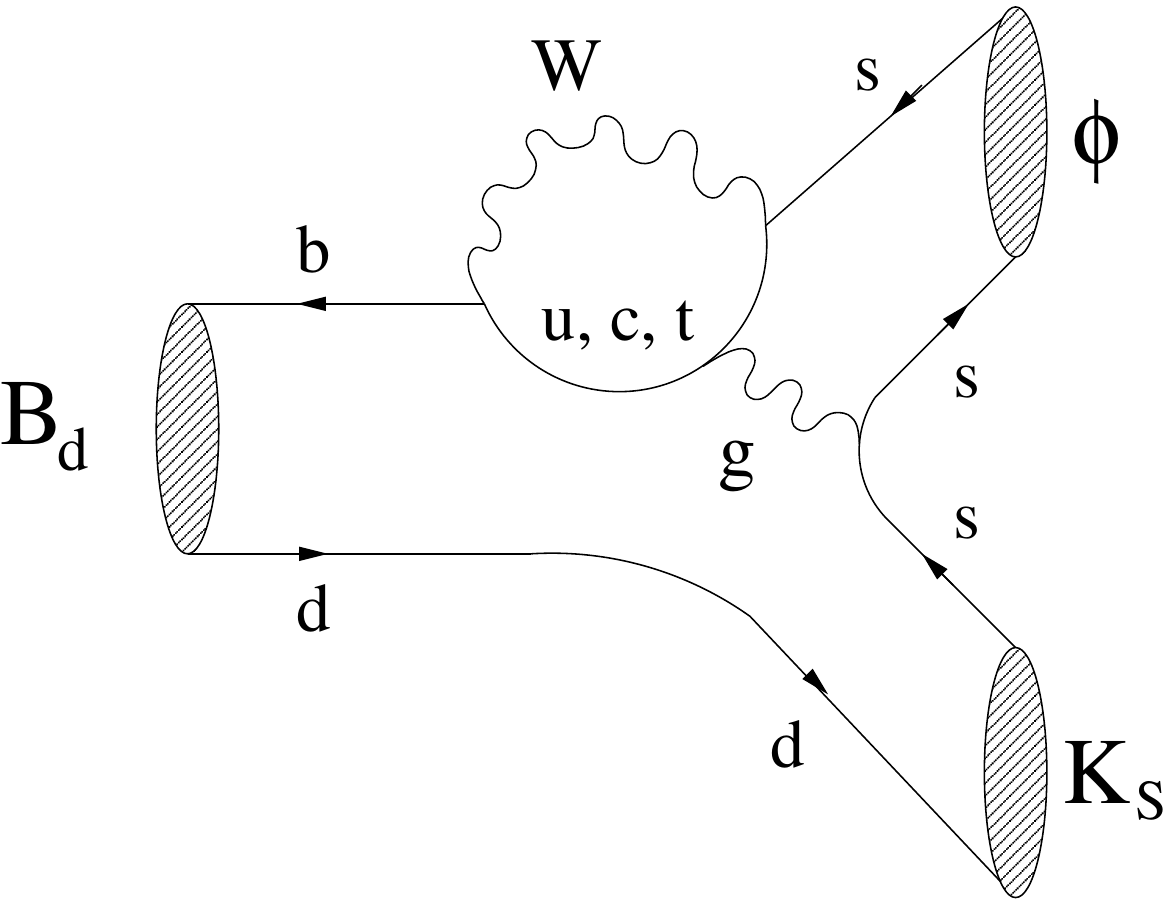}\hspace{1cm}
\includegraphics*[height=.2\textwidth]{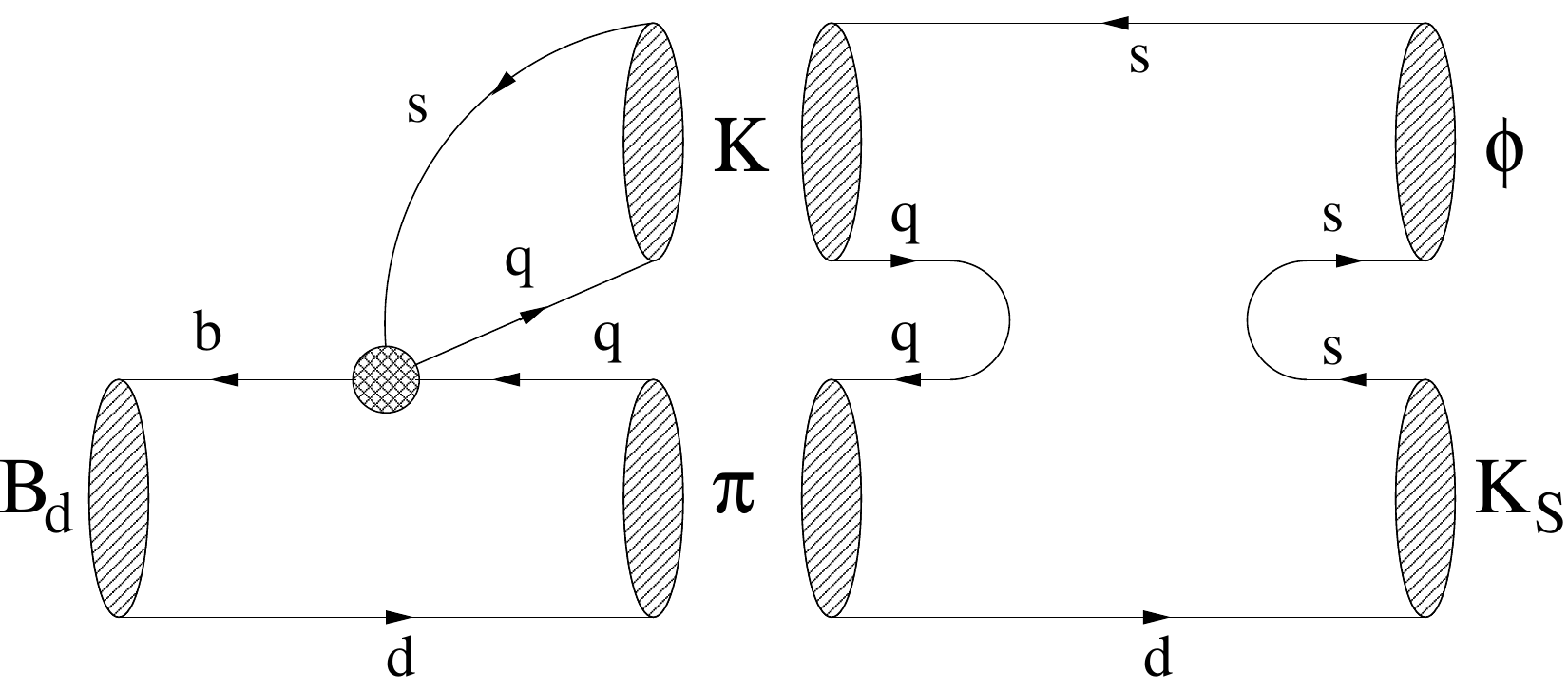}}
\caption{``Penguin" (left) and ``tree" (right) contributions to $B\to \phi
K_S$ (from Ref.~\cite{Fleischer:2002ys}).}
\label{fig:BphiK}
\end{figure}

Time dependent $CP$ violation in $b\to s$ dominated decays is a sensitive probe
of new physics.  Tree-level contributions to $b\to s\bar s s$ transitions are
expected to be small, and the penguin contributions to $B\to \phi K_S$ (left
diagram in Fig.~\ref{fig:BphiK}) are
\beq
\ov A_P = V_{cb} V_{cs}^*\, (P_c - P_t) + V_{ub} V_{us}^*\, (P_u - P_t)\,.
\eeq
Due to $|(V_{ub} V_{us}^*)/(V_{cb} V_{cs}^*)| \approx 0.02$ and expecting $|P_c
- P_t| / |P_u - P_t| = {\cal O}(1)$, the $B\to \phi K_S$ amplitude is also
dominated by a single weak phase, $V_{cb} V_{cs}^*$.  Therefore, the theory
uncertainty relating $S_{\phi K_S}$ to $\sin2\beta$ is small, although larger
than in $B\to \psi K_S$.  There is also a ``tree" contribution from $b\to
u\bar u s$ followed by $u\bar u \to s\bar s$ rescattering (right diagram in
Fig.~\ref{fig:BphiK}).  This amplitude is proportional to the suppressed CKM
combination, $V_{ub} V_{us}^*$, and it is actually not separable from
$P_u-P_t$.  Unless its matrix element is largely enhanced, it should not upset
the ${\rm Im} \lambda_{\phi K_S} = \sin2\beta + {\cal O}(\lambda^2)$ expectation
in the SM.  Similar reasons make many other modes, such as $B \to
\eta^{(\prime)} K_S$, $B_s \to \phi\phi$, etc., interesting and promising to
study.

\paragraph{The determinations of $\gamma$ and $\alpha$}

By virtue of Eq.~(\ref{angledef}), $\gamma$ does not depend on CKM elements
involving the top quark, so it can be measured in tree-level $B$ decays.  This
is an important distinction from $\alpha$ and $\beta$, and implies that $\gamma$
is less likely to be affected by BSM physics.

Most measurements of $\gamma$ utilize the fact that interference of $B^- \to D^0
K^-$ ($b\to c\bar u s$) and $B^-\to \D0bar K^-$ ($b \to u \bar c s$) transitions
can be studied in final states accessible in both $D^0$ and $\D0bar$
decays~\cite{Carter:1980hr}.  (A notable exception is the measurement from the
four time-dependent $\Bbar_s$ and $B_s \to D_s^\pm K^\mp$ rates, which is
possible at LHCb.)  It is possible to measure the $B$ and $D$ decay amplitudes,
their relative strong phases, and the weak phase $\gamma$ from the data.  There
are many variants, based on different $D$ decay channels~\cite{Gronau:1990ra,
Gronau:1991dp, Atwood:1996ci, Grossman:2002aq, bondar, Giri:2003ty}.  The best
current measurement comes from $D^0, \D0bar \to K_S\,\pi^+\pi^-$~\cite{bondar,
Giri:2003ty}, in which case both amplitudes are Cabibbo allowed, and the
analysis can be optimized by studying the Dalitz plot dependence of the
interference.  The world average of all $\gamma$ measurements
is~\cite{Hocker:2001xe}
\beq
\gamma = \big(73.2^{+6.3}_{-7.0}\big)^\circ.
\eeq
Most importantly, the theory uncertainty in the SM measurement is smaller
than the accuracy of any planned or imaginable future experiment.

The measurements usually referred to as determining $\alpha$, measure $\pi -
\beta - \gamma$, the third angle of the unitarity triangle in any model in which
the unitarity of the $3\times 3$ CKM matrix is maintained.  These measurements
are in time-dependent $CP$ asymmetries in $B\to\pi\pi$, $\rho\rho$, and
$\rho\pi$ decays.  In these decays the $b\to u\bar u d$ ``tree" amplitudes are
not much larger than the $b\to \sum_q q\bar q d$ ``penguin" contributions, which
have different weak phases.\footnote{Show that if the ``tree" amplitudes
dominated these decays then $\lambda_{\pi\pi}^{\rm (tree)} = e^{2i\alpha}$.} 
The tree contributions change isospin by $\Delta I = 3/2$ or 1/2, while the
penguin contribution is $\Delta I = 1/2$ only.  It is possible to use isospin
symmetry of the strong interaction to isolate $CP$ violation in the $\Delta
I=3/2$ channel, eliminating the penguin contributions~\cite{Gronau:1990ka,
Lipkin:1991st, Falk:2003uq}, yielding~\cite{Hocker:2001xe}
\beq
\alpha = \big(87.7^{+3.5}_{-3.3}\big)^\circ.
\eeq
Thus, the measurements of $\alpha$ are sensitive to new physics in
$B^0$\,--\,$\B0bar$ mixing and via possible $\Delta I = 3/2$ (or $\Delta I =
5/2$) contributions~\cite{Baek:2005cg}.

\paragraph{New physics in $B_d$ and $B_s$ mixing}

\begin{figure}[t]
\centerline{
\includegraphics[height=5cm,clip,bb=15 15 550 520]{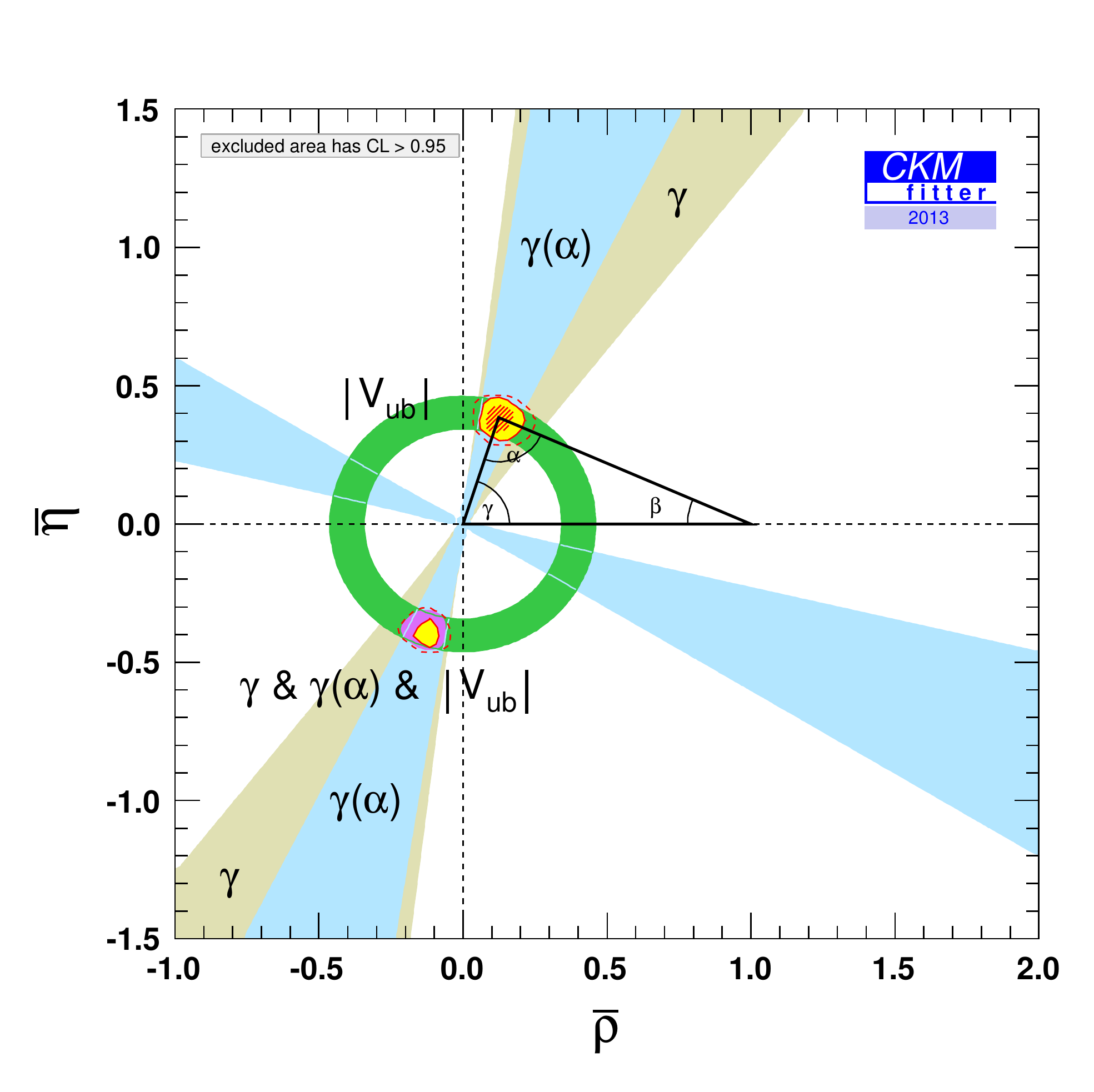}\hfill
\includegraphics[height=5cm,clip,bb=105 20 670 550]{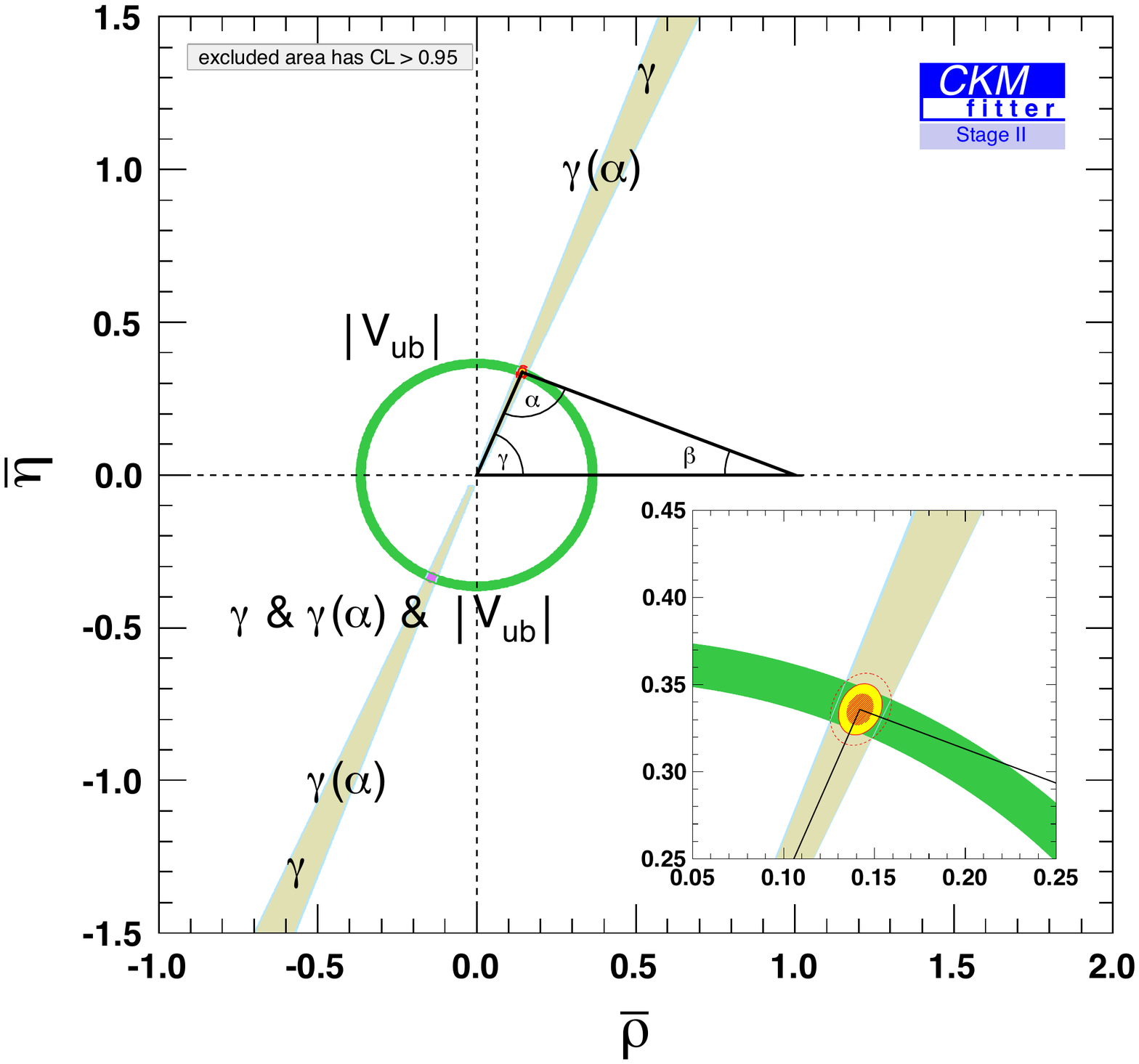}}
\caption{Constraints on $\rhobar - \etabar$, allowing new physics in the
$B_{d,s}$ mixing amplitudes.  Left plot shows the current constraints, right
plot is the expectation using 50\,ab$^{-1}$ Belle~II and 50\,fb$^{-1}$ LHCb
data.  Colored regions show 95\% CL, as in Fig.~\ref{fig:SMCKMfit}.
(From Ref.~\cite{Charles:2013aka}.)}
\label{fig:NPrhoeta}
\end{figure}

Although the SM CKM fit in Fig.~\ref{fig:SMCKMfit} shows impressive and
nontrivial consistency, the implications of the level of agreement are often
overstated.  Allowing new physics contributions, there are a larger number of
parameters related to $CP$ and flavor violation, and the fits become less
constraining.  This is shown in the left plot in Fig.~\ref{fig:NPrhoeta} where
the allowed region is indeed significantly larger than in
Fig.~\ref{fig:SMCKMfit} (the 95\% CL combined fit regions are indicated on
both plots).

\begin{figure}[tb]
\centerline{
\includegraphics[height=4.75cm,clip,bb=15 15 490 449]{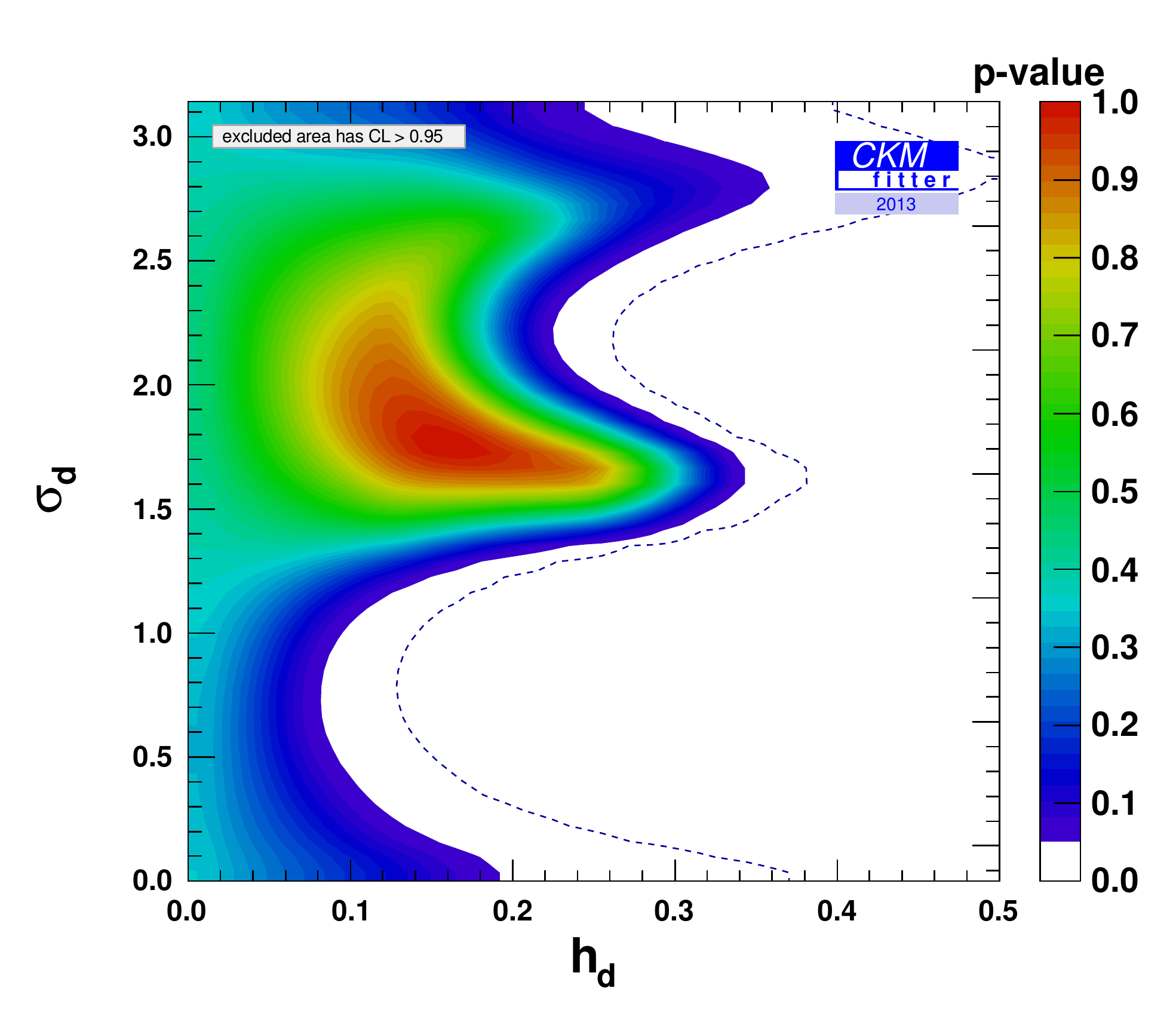}\hfill
\includegraphics[height=4.75cm,clip,bb=15 15 550 449]{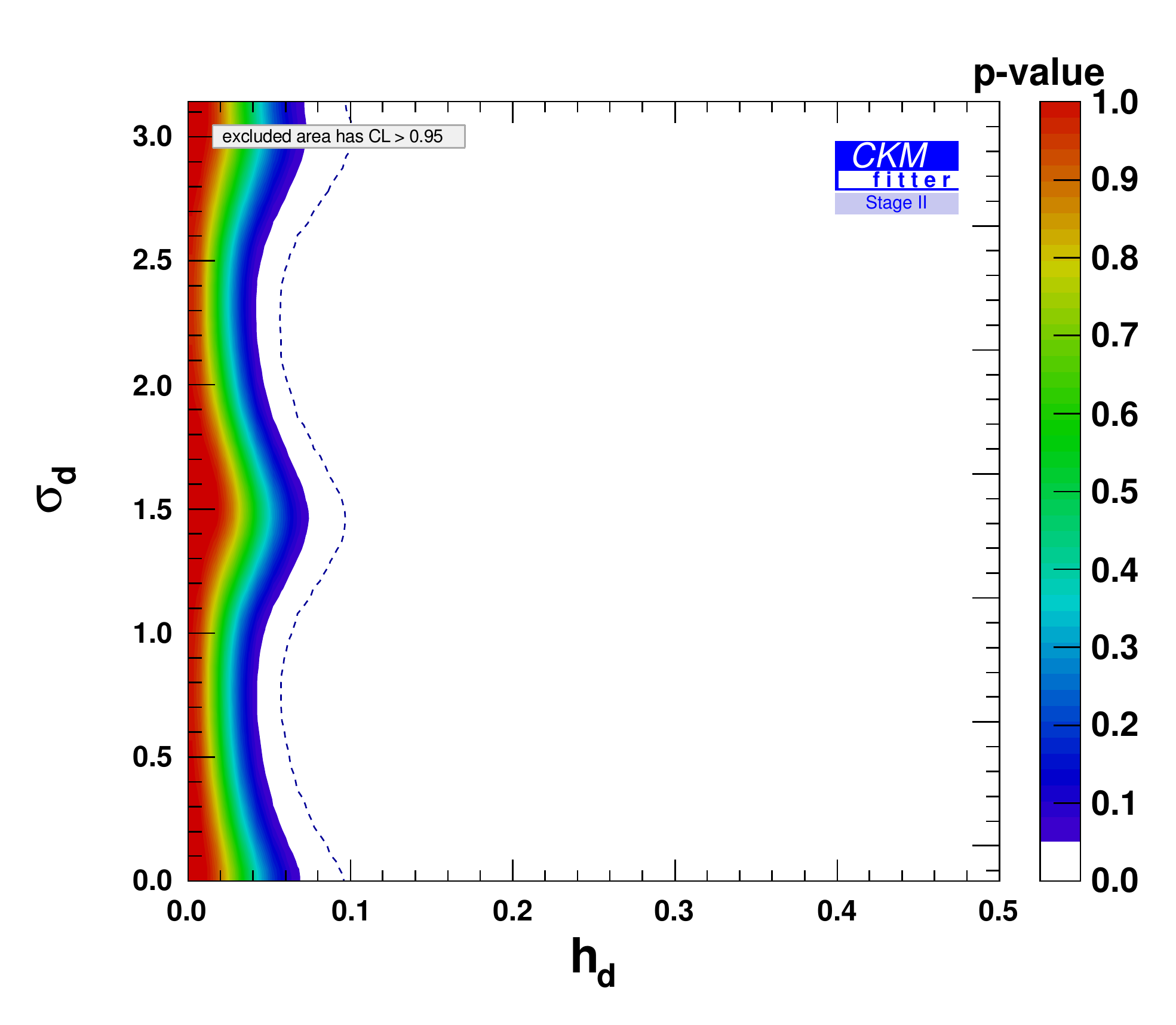}}
\vspace*{10pt}
\centerline{
\includegraphics[height=4.75cm,clip,bb=15 15 490 449]{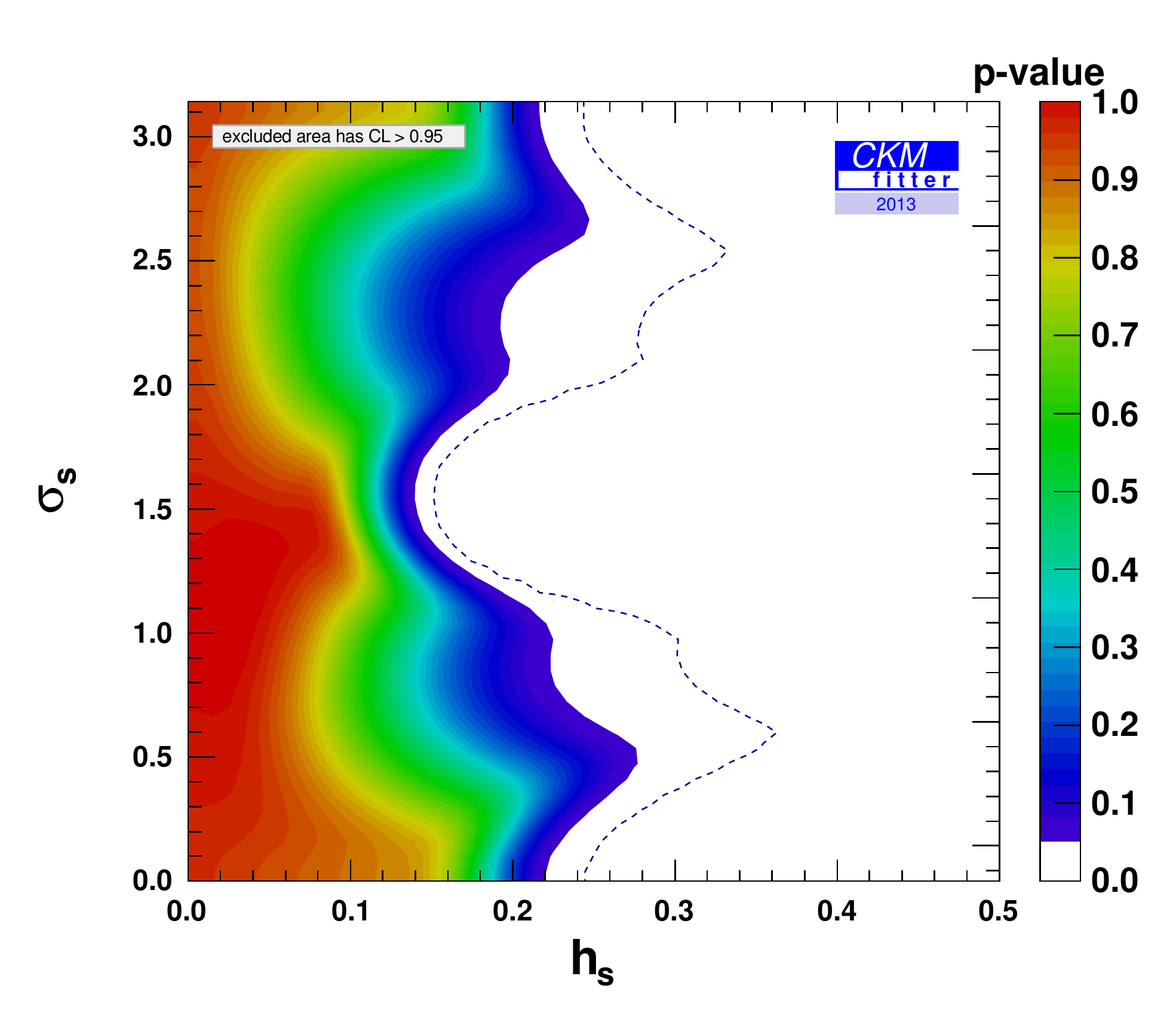}\hfill
\includegraphics[height=4.75cm,clip,bb=15 15 550 449]{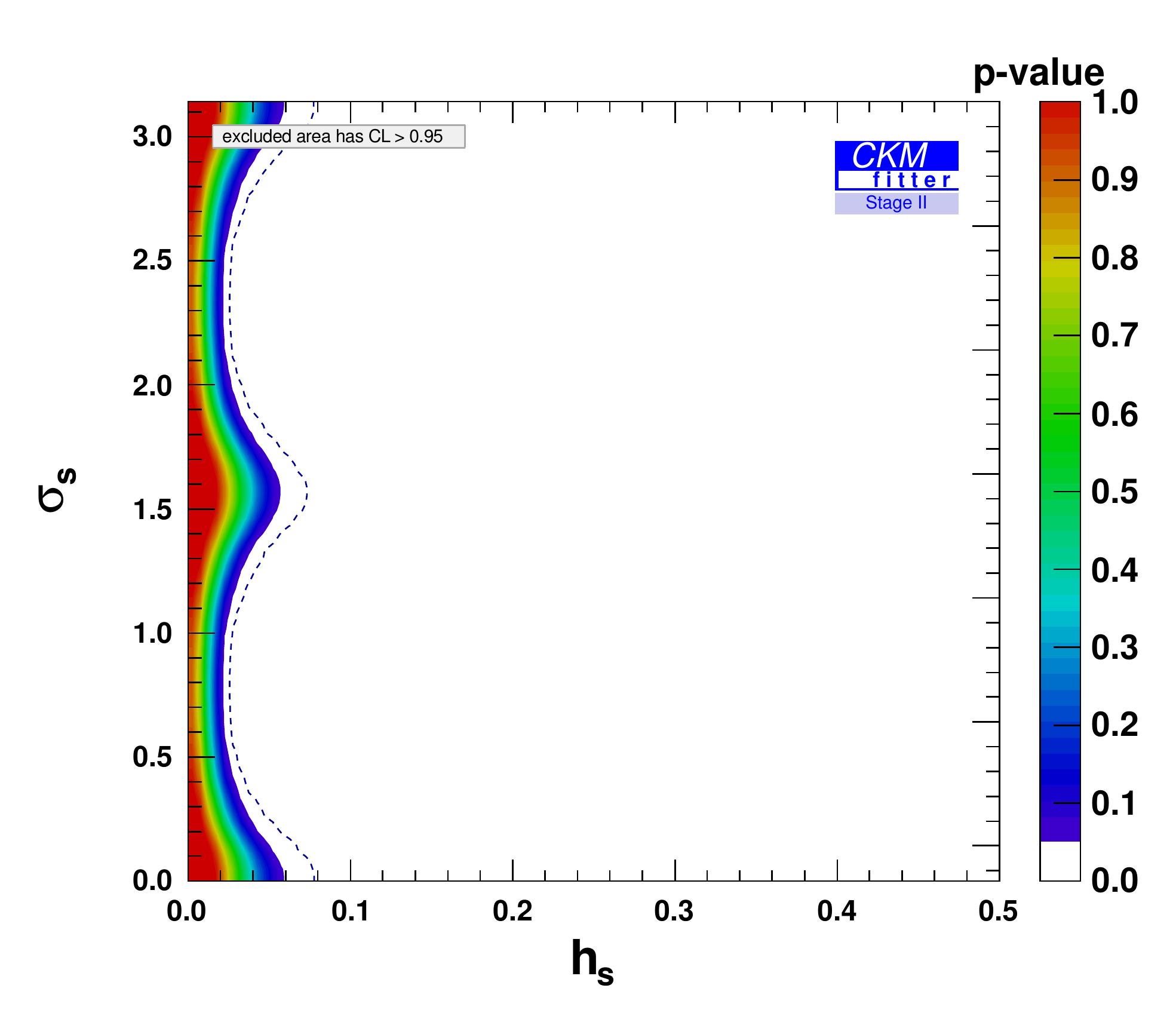}}
\caption{Constraints on the $h_d - \sigma_d$ (top row) and $h_s - \sigma_s$
parameters (bottom row).  Left plots show the current constraints, right plots
show those estimated to be achievable using 50\,ab$^{-1}$ Belle~II and
50\,fb$^{-1}$ LHCb data.  Colored regions show $2\sigma$ limits with the colors
indicating CL as shown, while the dashed lines show $3\sigma$ limits.  (From
Ref.~\cite{Charles:2013aka}.)}
\label{fig:NPmix}
\end{figure}

It has been known for decades that the mixing of neutral mesons is particularly
sensitive to new physics, and probes some of the highest scales.  In a large
class of models, NP has a negligible impact on tree-level SM transitions, and
the $3\times 3$ CKM matrix remains unitary.  (In such models $\alpha + \beta +
\gamma = \pi$ is maintained, and independent measurements of $\pi-\beta-\alpha$
and $\gamma$ can be averaged.)  We can parametrize the NP contributions to
neutral meson mixing as
\beq
M_{12} = M_{12}^{\rm SM} (1 + h_q\, e^{2i\sigma_q})\,, \qquad q=d,s\,.
\eeq
The constraints on $h_q$ and $\sigma_q$ in the $B_d^0$ and $B_s^0$ systems are
shown in the top and bottom rows of Fig.~\ref{fig:NPmix}, respectively.

For example, if NP modifies the SM operator describing $B$ mixing, by
\beq\label{SMmix}
\frac{C_{q}^2}{\Lambda^2}\, (\bar b_{L}\gamma^{\mu}q_{L})^2\,,
\eeq
then one finds
\beq
h_q \simeq \frac{|C_{q}|^2}{|V_{tb}^*\, V_{tq}|^2} 
  \left(\frac{4.5\, \TeV}{\Lambda}\right)^2\,.
\eeq
We can then translate the plots in Fig.~\ref{fig:NPmix} to the scale of
new physics probed.  The summary of expected sensitivities are shown in
Table~\ref{scaletable}. The sensitivities, even with SM-like loop- and
CKM-suppressed coefficients, are comparable to the scales probed by the LHC.

\begin{table}[t]\tabcolsep 14pt
\tbl{The scale of the operator in Eq.~(\ref{SMmix}) probed by $B_d^0$ and
$B_s^0$ mixings with 50\,ab$^{-1}$ Belle~II and 50\,fb$^{-1}$ LHCb data.  The
differences due to CKM-like hierarchy of couplings and/or loop suppression is
indicated. (From Ref.~\cite{Charles:2013aka}.)}
{\tablefont\begin{tabular}{c|c|c|c}
\hline\hline
\multirow{2}{*}{Couplings}  &  \raisebox{0pt}[9pt][0pt]{NP loop}  &  
  \multicolumn{2}{c}{Scales (TeV) probed by}\\
&  order  &  $B_d$ mixing  &  $B_s$ mixing   \\
\hline
$|C_{q}| = |V_{tb}V_{tq}^*|$ & tree level &  17  & 
  \raisebox{0pt}[9pt][0pt]{19} \\
\cdashline{2-4}
(CKM-like)  &  one loop  &  1.4  &  \raisebox{0pt}[9pt][0pt]{1.5}  \\
\hline
$|C_{q}| = 1$  &  tree level  &  $2\times 10^3$  &
  \raisebox{0pt}[9pt][0pt]{$5\times 10^2$}  \\
\cdashline{2-4}
(no hierarchy)  &  one loop &  $2\times 10^2$  &  \raisebox{0pt}[9pt][0pt]{40} \\
\hline\hline
\end{tabular}}
\label{scaletable}
\end{table}

\section{Some Implications of the Heavy Quark Limit}
\addcontentsline{toc}{}{Some Implications of the Heavy Quark Limit}

We have not directly discussed so far that most quark flavor physics processes
(other than top quark decays) involve strong interactions in a regime where
perturbation theory is not (or not necessarily) reliable.  The running of the
QCD coupling at lowest order is
\beq
\alpha_s(\mu) = {\alpha_s(\Lambda) \over \ds 
  1 + {\alpha_s\over 2\pi}\, \beta_0  \ln{\mu\over \Lambda}}\,,
\eeq
where $\beta_0 = 11 - 2 n_f/3$ and $n_f$ is the number of light quark flavors. 
Even in $B$ decays, the typical energy scale of certain processes can be a
fraction of $m_b$, possibly around or below a GeV.  The ways I know how to deal
with this in a tractable way are (i) symmetries of QCD, exact, or
approximate in some limits (CP invariance, heavy quark symmetry, chiral
symmetry); (ii)~the operator product expansion (for inclusive decays); (iii)
lattice QCD (for certain hadronic matrix elements).  An example of (i) is the
determination of $\sin2\beta$ from $B\to \psi K_S$, see Eq.~(\ref{BpsiKSCP}). 
So is the determination of $|V_{cb}|$ from $B\to D^*\ell\bar\nu$, see
Eq.~(\ref{F1}) below.  An example of (ii) is the analysis of inclusive $B\to
X_s\gamma$ decay rates discussed below, which provides some of the strongest
constraints on many TeV-scale BSM scenarios.

The role of (strong interaction) model-independent measurements cannot be
overstated.  To establish that a discrepancy between experiment and theory  is a
sign of new physics, model-independent predictions are crucial.  Results that
rely on modeling nonperturbative strong interaction effects will not disprove
the SM.  Most model-independent predictions are of the~form,
\beq
\mbox{Observable} = (\mbox{calculable terms}) \times 
  \bigg\{ 1 + \sum_{i,k} \big[\mbox{(small parameters)}_i\big]^k \bigg\} \,,
\eeq
where the small parameters can be $\lqcd/m_b$, $m_s/\Lambda_{\chi\rm SB}$,
$\alpha_s(m_b)$, etc.  For the purpose of these lectures, strong-interaction
model-independent means that the theoretical uncertainty is suppressed by small
parameters, so that theorists argue about ${\cal O}(1)\times$(small numbers)
instead of ${\cal O}(1)$ effects.  There are always theoretical uncertainties
suppressed by some $(\mbox{small parameter})^n$, which cannot be calculated from
first principles.  If the goal is to test the SM, one must assign ${\cal O}(1)$
uncertainties in such terms.

In addition, besides formal suppressions of certain corrections in some limits,
experimental guidance is always needed to establish how well an expansion works;
for example, $f_\pi$, $m_\rho$, and $m_K^2/m_s$ are all of order $\lqcd$, but
their numerical values span an order of magnitude.

\paragraph{Heavy quark symmetry (HQS)}

In hadrons composed of heavy quarks the dynamics of QCD simplifies.  Mesons
containing a heavy quark -- heavy antiquark pair, $Q\ov Q$, form
positronium-type bound states, which become perturbative in the limit $m_Q \gg
\lqcd$~\cite{Appelquist:1974zd}.  In mesons composed of a heavy quark, $Q$, and
a light antiquark, $\bar q$ (and gluons and $q\bar q$ pairs), the heavy quark
acts as a static color source with fixed four-velocity, $v^\mu$, and the wave
function of the light degrees of freedom (the ``brown muck") become insensitive
to the spin and mass (flavor) of the heavy quark, resulting in heavy quark
spin-flavor symmetries~\cite{Isgur:1989vq}.

The physical picture is similar to atomic physics, where simplifications occur
due to the fact that the electron mass, $m_e$, is much smaller than the nucleon
mass, $m_N$.  The analog of flavor symmetry is that isotopes have similar
chemistry, because the electrons' wave functions become independent of $m_N$ in
the $m_N \gg m_e$ limit.  The analog of spin symmetry is that hyperfine levels
are almost degenerate, because the interaction of the electron and nucleon spin
diminishes in the $m_N \gg m_e$ limit.

\paragraph{Spectroscopy of heavy-light mesons}

The spectroscopy of heavy hadrons simplifies due to heavy quark symmetry.  We
can write the angular momentum of a heavy-light meson as $J = \vec s_Q + \vec
s_l$, where $\vec s_l$ is the total angular momentum of the light degrees of
freedom.  Angular momentum conservation, $[\vec J, {\cal H}] = 0$, and heavy
quark symmetry, $[\vec s_Q, {\cal H}] = 0$, imply $[\vec s_l, {\cal H}] = 0$. 
In the $m_Q \gg \lqcd$ limit, the spin of the heavy quark and the total angular
momentum of light degrees of freedom are separately conserved, modified only by
subleading interactions suppressed by $\lqcd/m_Q$.

Thus, hadrons containing a single heavy quark can be labeled with $s_l$, and for
any value of $s_l$ there are two (almost) degenerate states with total angular
momentum $J_\pm = s_l \pm \frac12$.  (An exception occurs for the lightest
baryons containing a heavy quark, when $s_l = 0$, and there is a single state
with $J = \frac12$, the $\Lambda_b$ and $\Lambda_c$.)  The ground state mesons
with $Q\bar q$ flavor quantum numbers contain light degrees of freedom with
spin-parity $s_l^{\pi_l} = \frac12^-$, giving a doublet containing a spin zero
and spin one meson.  For $Q=c$ these are the $D$ and $D^*$, while $Q=b$ gives
the $B$ and $B^*$ mesons.

\begin{figure}[t]
\centerline{\includegraphics*[width=.5\textwidth]{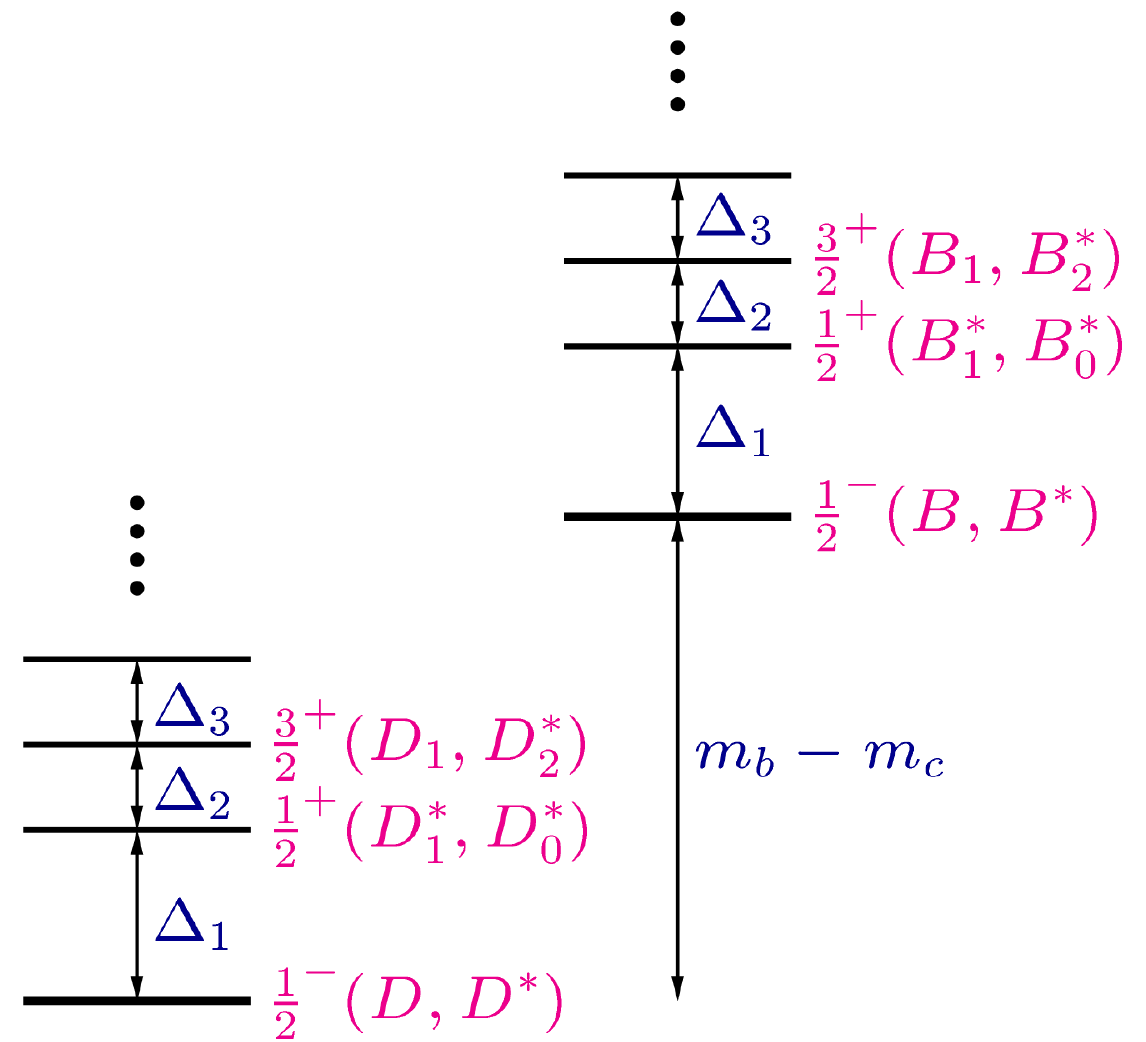}}
\caption{Spectroscopy of $B$ and $D$ mesons.  For each doublet level, the
spin-parity of the light degrees of freedom, $s_l^{\pi_l}$, and the names of the
physical states are indicated.}
\label{fig:spectra}
\end{figure}

The mass splittings between the doublets, $\Delta_i$, are of order $\lqcd$, and
are the same in the $B$ and $D$ sectors at leading order in $\lqcd/m_Q$, as
illustrated in Fig.~\ref{fig:spectra}.  The mass splittings within each doublet
are of order $\lqcd^2/m_Q$.  This is supported by experimental data; e.g., for
the $s_l^{\pi_l} = \frac12^-$ ground state doublets $m_{D^*}-m_D \approx
140\,$MeV while $m_{B^*}-m_B \approx 45\,$MeV, and their ratio, 0.3, is
consistent with $m_c/m_b$.

Let us mention a puzzle.  The mass splitting of the lightest vector and
pseudoscalar mesons being ${\cal O}(\lqcd^2/m_Q)$ implies that $m_V^2-m_P^2$ is
approximately constant.  This argument relies on $m_Q \gg \lqcd$.  The data are
\beq
\begin{array}{rclrcl}
m_{B^*}^2 - m_B^2 &=& 0.49\,\GeV^2\,,\qquad  &
  m_{B_s^*}^2 - m_{B_s}^2 &=& 0.50\,\GeV^2\,, \\
m_{D^*}^2 - m_D^2 &=& 0.54\,\GeV^2\,,  &
  m_{D_s^*}^2 - m_{D_s}^2 &=& 0.58\,\GeV^2\,, \\
m_\rho^2 - m_\pi^2 &=& 0.57\,\GeV^2\,,  &
  m_{K^*}^2 - m_K^2 &=& 0.55\,\GeV^2\,.
\end{array}
\eeq
It is not understood why the light meson mass splittings in the last line are
so close numerically.  (It is expected in the nonrelativistic constituent quark
model, which fails to account for several properties of these mesons.)  There
must be something more going on than heavy quark symmetry, and if this were its
only prediction, we could not say that there is strong evidence that it is
useful.  So in general, to understand a theory, it is not only important how
well it works, but also how it breaks down outside its range of validity.

\paragraph{Heavy quark effective theory (HQET)}

The consequences of heavy quark symmetry and the corrections to the symmetry
limit can be studied by constructing an effective theory which makes the
consequences of heavy quark symmetry explicit.  The heavy quark in a heavy-light
meson is almost on-shell, so we can expand its momentum as $p_Q^\mu = m_Q v^\mu
+ k^\mu$, where $|k| = {\cal O}(\lqcd)$ and $v^2=1$.  Expanding the heavy quark
propagator,
\beq
{i\over\psla-m_Q} = {i(\psla+m_Q)\over p^2-m_Q^2}
  = {i(m_Q\vsla+\ksla+m_Q)\over 2m_Q\,v\cdot k +k^2}
  = {i\over v\cdot k}\,{1+\vsla\over2}+ \ldots\,.
\eeq
it becomes independent of the heavy quark mass, a manifestation of heavy quark
flavor symmetry.  Hence the Feynamn rules simplify,
\def\propline{\hrule width60pt depth0.5pt height0.5pt}
\def\prop{\vcenter{\propline}}
\def\heavyprop{\vcenter{\propline\kern1.0pt\propline}}
\newdimen\unit
\def\point#1 #2 #3{\rlap{\kern#1\unit
  \raise#2\unit\hbox to 0pt{\hss$#3$\hss}}}
\def\cpt#1 #2 {\point #1 #2 \cdot}
\beq\label{propagator}
\vspace{2pt}
\begin{matrix}
  \prop & \quad \longrightarrow \quad & \heavyprop \\[4pt]
  \ds {i\over \psla -m_Q} && \ds {i\over v\cdot k}\, P_+(v)\,,
\end{matrix}
\eeq
where $P_\pm = (1 \pm \vsla)/2$ are projection operators, and the double line
denotes the heavy quark propagator.  In the rest frame of the heavy quark, $P_+
= (1+\gamma^0)/2$ projects onto the heavy quark (rather than anti-quark)
components.  The coupling of a heavy quark to gluons simplifies due to
\beq
P_+\gamma^\mu P_+ = P_+ v^\mu P_+ = v^\mu P_+ \,,
\eeq
hence we can replace
\newbox\cycloid
\setbox\cycloid=\hbox{\unit=0.5pt
\cpt 0.0000 6.0000 \cpt 0.4998 5.9896 \cpt 0.9986 5.9585 \cpt
1.4954 5.9067 \cpt 1.9895 5.8343 \cpt 2.4799 5.7412 \cpt 2.9658
5.6277 \cpt 3.4463 5.4938 \cpt 3.9205 5.3396 \cpt 4.3874 5.1653
\cpt 4.8461 4.9709 \cpt 5.2957 4.7567 \cpt 5.7351 4.5228 \cpt
6.1633 4.2695 \cpt 6.5792 3.9969 \cpt 6.9819 3.7055 \cpt 7.3700
3.3954 \cpt 7.7424 3.0671 \cpt 8.0979 2.7209 \cpt 8.4351 2.3573
\cpt 8.7526 1.9769 \cpt 9.0489 1.5802 \cpt 9.3225 1.1679 \cpt
9.5716 0.7409 \cpt 9.7944 0.3001 \cpt 9.9889 -0.1533 \cpt 10.1530
-0.6180 \cpt 10.2844 -1.0924 \cpt 10.3805 -1.5744 \cpt 10.4386
-2.0619 \cpt 10.4555 -2.5516 \cpt 10.4281 -3.0401 \cpt 10.3528
-3.5226 \cpt 10.2257 -3.9932 \cpt 10.0433 -4.4442 \cpt 9.8020
-4.8659 \cpt 9.4994 -5.2457 \cpt 9.1354 -5.5677 \cpt 8.7141
-5.8131 \cpt 8.2464 -5.9621 \cpt 7.7516 -5.9974 \cpt 7.2570
-5.9114 \cpt 6.7910 -5.7096 \cpt 6.3766 -5.4093 \cpt 6.0267
-5.0329 \cpt 5.7457 -4.6020 \cpt 5.5322 -4.1342 \cpt 5.3823
-3.6432 \cpt 5.2912 -3.1389 \cpt 5.2537 -2.6287 \cpt 5.2653
-2.1183 \cpt 5.3216 -1.6118 \cpt 5.4188 -1.1123 \cpt 5.5536
-0.6224 \cpt 5.7227 -0.1439 \cpt 5.9236 0.3214 \cpt 6.1536 0.7724
\cpt 6.4107 1.2080 \cpt 6.6926 1.6274 \cpt 6.9975 2.0298 \cpt
7.3238 2.4147 \cpt 7.6697 2.7814 \cpt 8.0338 3.1297 \cpt 8.4146
3.4591 \cpt 8.8108 3.7693 \cpt 9.2213 4.0600 \cpt 9.6446 4.3310
\cpt 10.0798 4.5820 \cpt 10.5257 4.8130 \cpt 10.9813 5.0238 \cpt
11.4455 5.2142 \cpt 11.9174 5.3841 \cpt 12.3960 5.5335 \cpt
12.8803 5.6622 \cpt 13.3694 5.7702 \cpt 13.8625 5.8576 \cpt
14.3587 5.9241 \cpt 14.8569 5.9699 \cpt 15.3565 5.9949 }
\def\gluon{\vcenter{\hbox{\copy\cycloid}}\hbox to 7.85pt{\hfil}}
\beq
\begin{matrix}
\gluon\gluon\gluon\gluon \vcenter{\hrule height 40pt width 1pt}
  & \quad\longrightarrow\quad &
  \gluon\gluon\gluon\gluon \vcenter{\hbox{\vrule height 40pt width 1pt
  \kern0.8pt \vrule height40pt width1pt}} \\
\noalign{\vskip7pt}
\ds ig \gamma^\mu\, {\lambda^a\over2} && \ds ig v^\mu\, {\lambda^a\over2}\,.\\
\end{matrix}
\eeq
The lack of any $\gamma$ matrix is a manifestation of heavy quark spin symmetry.

To derive the effective Lagrangian of HQET, it is convenient to decompose the
four-component Dirac spinor as
\beq
Q(x) = e^{-i m_Q v\cdot x} \big[Q_v(x) + {\cal Q}_v(x)\big]\,,
\eeq
where
\beq\label{hqfield}
Q_v(x) = e^{i m_Q v\cdot x}\, P_+(v)\, Q(x) \,, \qquad
{\cal Q}_v(x) = e^{i m_Q v\cdot x}\, P_-(v)\, Q(x) \,.
\eeq
The $e^{i m_Q v\cdot x}$ factor subtracts $m_Q v$ from the heavy quark
momentum.  At leading order only $Q_v$ contributes, and the effects of ${\cal
Q}_v$ are suppressed by powers of $\lqcd/m_Q$.  The heavy quark velocity, $v$,
acts as a label of the heavy quark fields~\cite{Georgi:1990um}, because $v$
cannot be changed by soft interactions.  In terms of these fields the QCD
Lagrangian simplifies,
\beq\label{Lag}
{\cal L} = \bar Q(i \Dsla - m_Q) Q = \bar Q_v i \Dsla Q_v + \ldots
  =  \bar Q_v (i v\cdot D) Q_v + \ldots \,, 
\eeq
where the ellipses denote terms suppressed by powers of $\lqcd/m_Q$.  The
absence of any Dirac matrix is a consequence of heavy quark symmetry, which
implies that the heavy quark's propagator and its coupling to gluons are
independent of the heavy quark spin.  This effective theory provides a framework
to calculate perturbative ${\cal O}(\alpha_s)$ corrections  and to parametrize
nonperturbative ${\cal O}(\lqcd/m_Q)$ terms.

\paragraph{Semileptonic $B\to D^{(*)}\ell\bar\nu$ decays and $|V_{cb}|$}

Heavy quark symmetry is particularly predictive for these decays.  In the
$m_{b,c} \gg \lqcd$ limit, the configuration of the brown muck only depends on
the four-velocity of the heavy quark, but not on its mass and spin.  So when the
weak current changes suddenly (on a time scale $\ll \lqcd^{-1}$) the flavor
$b\to c$, the momentum $\vec p_b \to \vec p_c$, and possibly flips the spin,
$\vec s_b\to \vec s_c$, the brown muck only feels that the four-velocity of the
static color source changed, $v_b \to v_c$.  Therefore, the matrix elements that
describe the transition probabilities from the initial to the final state are
independent of the Dirac structure of weak current, and can only depend on a
scalar quantity, $w \equiv v_b \cdot v_c$. 

The ground-state pseudoscalar and vector mesons for each heavy quark flavor (the
spin symmetry doublets $D^{(*)}$ and $B^{(*)}$) can be represented by a
``superfield", combining fields with different spins, that has the right
transformation property under heavy quark and Lorentz symmetry,
\beq\label{superfield}
{\cal M}_v^{(Q)} = \frac{1+\vsla}2 \Big[ \gamma^\mu M_\mu^{*(Q)}(v,\varepsilon)
- i\gamma_5 M^{(Q)}(v) \Big] .
\eeq
The $B^{(*)} \to D^{(*)}$ matrix element of any current can be parametrized as
\beq\label{traceformula}
\langle M^{(c)}(v') \,|\, \bar c_{v'}\, \Gamma\, b_v \,|\, M^{(b)}(v) \rangle
  = {\rm Tr}\, \Big[ F(v,v')\, \bar{\cal M}_{v'}^{(c)}\, \Gamma\, 
  {\cal M}_v^{(b)} \Big] .
\eeq
Because of heavy quark symmetry, there cannot be other Dirac matrices
between the $\bar{\cal M}_{v'}^{(c)}$ and ${\cal M}_v^{(b)}$ fields.  The most
general form of $F$ is
\beq
F(v,v') = f_1(w) + f_2(w) \vsla + f_3(w) \vsla' + f_4(w) \vsla\vsla' .
\eeq
As stated above, $w \equiv v \cdot v'$ is the only possible scalar, simply
related to  $q^2 = (p_B - p_{D^{(*)}})^2 = m_B^2 + m_{D^{(*)}}^2 - 2m_B
m_{D^{(*)}} w$.  Using ${\cal M}_v^{(Q)} =  P_+(v) {\cal M}_v^{(Q)} P_-(v)$, we
can write
\beqa
F & \doteq & P_-(v) F P_-(v') = \big[ f_1(w)-f_2(w)-f_3(w)+f_4(w)\big] P_-(v) P_-(v') \nn\\
  &=& \xi(w)\, P_-(v)\, P_-(v') \;\doteq\; \xi(w) \,.
\eeqa
This defines the Isgur-Wise function, $\xi(w)$, and $\doteq$ denotes relations
valid when evaluated inside the trace in Eq.~(\ref{traceformula}).

\begin{figure}[t]
\centerline{\includegraphics*[width=.5\textwidth]{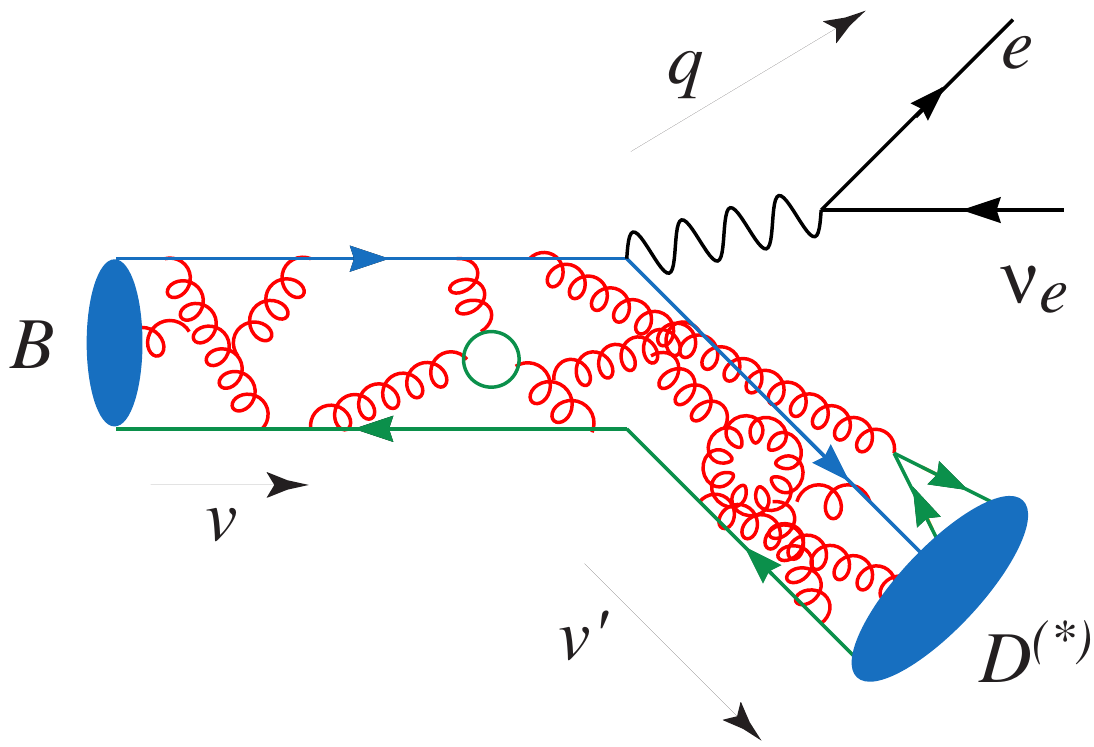}}
\caption{Illustration of strong interactions parametrized by the
Isgur-Wise function.}
\label{fig:BDlnu}
\end{figure}

Since only weak interactions change $b$-quark number, the matrix element of
$\bar b\gamma_0 b$, the $b$-quark number current, is $\langle B(v)|\bar
b\gamma_0 b|B(v)\rangle = 2m_B v_0$.  Comparing it with the result obtained
using Eq.~(\ref{traceformula}),
\beq\label{IWnorm}
\langle B(v)|\bar b\gamma_\mu b|B(v)\rangle = 2m_B v_\mu\, \xi(1)\,,
\eeq
implies that $\xi(1) = 1$.  That is, at $w=1$, the ``zero recoil" point, when
the $D^{(*)}$ is at rest in the rest-frame of the decaying $B$ meson,  the
configuration of the brown muck does not change at all, and heavy quark symmetry
determines the hadronic matrix element (see Fig.~\ref{fig:BDlnu}).  Moreover,
the six form factors that describe semileptonic $B\to D^{(*)}\ell\bar\nu$ decays
are related to this universal function, which contains all the low energy
nonperturbative hadronic physics relevant for these decays.\footnote{Using only
Lorentz invariance, six form factors parametrize $B\to D^{(*)} \ell \bar\nu$
decay,
\beqa\label{BDffdef}
\langle D(v') | V_\nu | B(v)\rangle &=& \sqrt{m_Bm_D}\, \big[ h_+\,(v+v')_\nu
  + h_-\, (v-v')_\nu \big] , \nn\\
\langle D^*(v') | V_\nu | B(v)\rangle &=& i \sqrt{m_Bm_{D^*}}\, h_V\,
  \epsilon_{\nu\alpha\beta\gamma}\epsilon^{*\alpha}v'^\beta v^\gamma , \nn\\
\langle D(v') | A_\nu | B(v)\rangle &=& 0 , \\
\langle D^*(v') | A_\nu | B(v)\rangle &=& \sqrt{m_Bm_{D^*}}\,
  \big[h_{A_1}\,(w+1)\epsilon^*_\nu - h_{A_2}\,(\epsilon^*\cdot v)v_\nu
  - h_{A_3}\,(\epsilon^*\cdot v)v'_\nu\big] , \nn
\eeqa
where $V_\nu = \bar c\gamma_\nu b$, $A_\nu = \bar c\gamma_\nu\gamma_5 b$, and
$h_i$ are functions of $w$.  Show that this is indeed the most general form of
these matrix elements, and that at leading order in $\lqcd/m_Q$,
\beq
h_+(w) = h_V(w) = h_{A_1}(w) = h_{A_3}(w) = \xi(w)\,, \qquad
  h_-(w) = h_{A_2}(w) = 0\,.
\eeq}

The determination of $|V_{cb}|$ from $B\to D^{(*)} \ell \bar\nu$ decays use
fits to the decay distributions to measure the rates near zero recoil, $w=1$.
The rates can be schematically written as
\begin{equation}\label{rates}
{\d\Gamma(B\to D^{(*)} \ell\bar\nu)\over \d w} 
  = (\mbox{calculable})\, |V_{cb}|^2
\begin{cases}
(w^2-1)^{1/2}\, {\cal F}_*^2(w), & \mbox{for } B\to D^*, \\
(w^2-1)^{3/2}\, {\cal F}^2(w), & \mbox{for } B\to D\,. 
\end{cases}
\end{equation}
Both ${\cal F}(w)$ and ${\cal F}_*(w)$ are equal to the Isgur-Wise function in
the $m_Q \to\infty$ limit, and ${\cal F}_{(*)}(1) = 1$ is the basis for a
model-independent determination of $|V_{cb}|$.  There are calculable 
corrections in powers of $\alpha_s(m_{c,b})$, as well as terms suppressed by
$\lqcd/m_{c,b}$, which can only be parametrized, and that is where hadronic
uncertainties enter.  Schematically,
\beqa\label{F1}
{\cal F}_*(1) &=& 1_{\mbox{\footnotesize (Isgur-Wise)}} + c_A(\alpha_s) 
  + {0_{\mbox{\footnotesize (Luke)}}\over m_{c,b}} 
  + {(\mbox{lattice or models})\over m_{c,b}^2} + \ldots \,, \nn\\
{\cal F}(1) &=& 1_{\mbox{\footnotesize (Isgur-Wise)}} + c_V(\alpha_s) 
  + {(\mbox{lattice or models})\over m_{c,b}} + \ldots \,.
\eeqa
The absence of the ${\cal O}(\lqcd/m_{c,b})$ term for $B\to D^{*}\ell\bar\nu$ at
zero recoil is a consequence of Luke's theorem~\cite{Luke:1990eg}.  Calculating
corrections to the heavy quark limit in these decays is a vast subject.  Heavy
quark symmetry also makes model-independent predictions for $B$ decays to
excited $D$ mesons~\cite{Leibovich:1997tu}.  It is due to heavy quark symmetry
that the SM predictions for the recently observed anomalies in the $B\to
D^{(*)}\tau\bar\nu$ branching ratios~\cite{Lees:2012xj} are under good
theoretical control.

\paragraph{Inclusive semileptonic decays and $B\to X_s\gamma$}

Instead of identifying all final-state particles in a decay, sometimes it is
useful to sum over final-state hadrons that can be produced by the strong
interaction, subject to constraints determined by short-distance physics, e.g.,
the energy of a photon or a charged lepton.  Although hadronization is
nonperturbative, it occurs on much longer distance (and time) scales than the
underlying weak decay.  Typically we are interested in a quark-level transition,
such as $b\to c\ell\bar\nu$, $b\to s\gamma$, etc., and we would like to extract
from the data short distance parameters, $|V_{cb}|$, $C_7(m_b)$, etc.  To do
this, we need to relate the quark-level operators to the measurable decay rates.

For example, consider inclusive semileptonic $b\to c$ decay mediated by
\beq
O_{\rm sl} = -{4G_F\over \sqrt2}\, V_{cb}\, 
  (J_{bc})^\alpha\, (J_{\ell\nu})_\alpha \,,
\eeq
where $J^\alpha_{bc} = \bar c\, \gamma^\alpha P_L b$ and
$J^\beta_{\ell\nu} = \bar\ell\, \gamma^\beta P_L \nu$.
The decay rate is given by the square of the matrix element, integrated over
phase space, and summed over final states,
\beq
\Gamma(B\to X_c\ell\bar\nu) \sim \sum_{X_c} \int \d [{\rm PS}]\,
  \big| \langle X_c\ell\bar\nu | O_{\rm sl} | B \rangle \big|^2 .
\eeq
Since leptons have no strong interaction, the squared matrix element and phase
space factorize into $B\to X_c W^*$ and a perturbatively calculable leptonic
part, $W^* \to \ell\bar\nu$.  The nontrivial part is the hadronic tensor,
\beqa
W^{\mu\nu} &=& \sum_{X_c} (2\pi)^3\, \delta^4(p_B-q-p_X)\,
  \langle B | J^{\mu\dagger}_{bc} | X_c \rangle\, 
  \langle X_c | J^\nu_{bc} | B \rangle \nn\\*
&=& \frac1\pi\, {\rm Im} \int\! \d^4 x\, e^{-iq\cdot x}\,
  \langle B |\, T \big\{ J^{\mu\dagger}_{bc}(x)\, 
  J^\nu_{bc}(0) \big\}\, | B \rangle \,,
\eeqa
where the second line is obtained using the optical theorem, and $T$ denotes
here the time-ordered product of the operators.  It is this time-ordered product
that can be expanded in an operator product expansion (OPE)~\cite{Chay:1990da,
Bigi:1992su, Bigi:1993fe, Manohar:1993qn}.  In the $m_b \gg \lqcd$ limit, the
time-ordered product is dominated by short distances, $x \ll \lqcd^{-1}$, and
one can express the hadronic tensor $W^{\mu\nu}$ as a sum of matrix elements of
local operators.  Schematically,
\beqa\label{opesketch}
\raisebox{-28pt}{\includegraphics*[width=.32\textwidth]{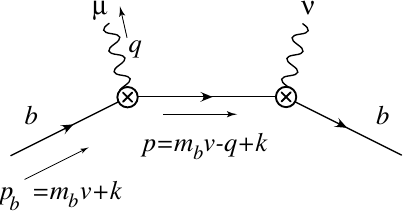}} &=& 
  \raisebox{-16pt}{\includegraphics*[width=.12\textwidth]{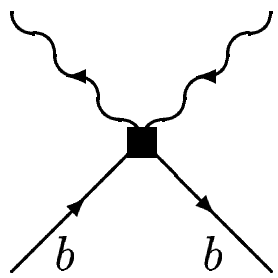}}
 \!+ {0\over m_b}\! \raisebox{-16pt}{\includegraphics*[width=.12\textwidth]{ope2}}
 \!+ {1\over m_b^2}\! \raisebox{-16pt}{\includegraphics*[width=.12\textwidth]{ope2}}
 \!+ \ldots\qquad
\eeqa
This is analogous to the multipole expansion.  At leading order in $\lqcd/m_b$
the lowest dimension operator is $\bar b\, \Gamma\, b$, where $\Gamma$ is some
(process-dependent) Dirac matrix.  Its matrix element is determined by the $b$
quark content of the initial state using Eqs.~(\ref{traceformula}) and
(\ref{IWnorm}); therefore, inclusive $B$ decay rates in the $m_b \gg \lqcd$
limit are equal to the $b$ quark decay rates.  Subleading effects are
parametrized by matrix elements of operators with increasing number of
derivatives, which are sensitive to the distribution of chromomagnetic and
chromoelectric fields.  There are no ${\cal O}(\lqcd/m_b)$ corrections, because
the $B$ meson matrix element of any dimension-4 operator vanishes, $\langle B(v)
|\, \bar Q_v^{(b)}\, iD_\alpha \Gamma\, Q_v^{(b)} |B(v) \rangle = 0$.  The
leading nonperturbative effects, suppressed by $\lqcd^2/m_b^2$, are
parametrized by two HQET matrix elements, denoted by $\lambda_{1,2}$.  This is
the basis of the model-independent determinations of $m_b$ and $|V_{cb}|$ from
inclusive semileptonic $B$ decays.

Some important applications, such as $B\to X_s\gamma$~\cite{Misiak:2006zs}  or
$B\to X_u\ell\bar\nu$, are more complicated.  Near boundaries of phase space,
the energy release to the hadronic final state may not be large.  One can think
of the OPE as an expansion in the residual momentum of the $b$ quark, $k$, shown
in Eq.~(\ref{opesketch}),
\beq
{1\over (m_b v + k -q)^2-m_q^2} 
  = {1\over [(m_b v -q)^2-m_q^2] + [2k\cdot (m_b v-q)] + k^2} \,.
\eeq
For the expansion in $k$ to converge, the final state phase space can
only be restricted in a way that allows hadronic final states, $X$, to
contribute with
\beq\label{converge}
m_X^2-m_q^2 \gg E_X \lqcd \gg \lqcd^2 \,.
\eeq
In $B\to X_s\gamma$ when an experimental lower cut is imposed on $E_\gamma$ to
reject backgrounds, the left-most inequality can be violated.  The same occurs
in $B\to X_u\ell\bar\nu$ when experimental cuts are used to suppress $B\to
X_c\ell\bar\nu$ backgrounds.  If the right-most inequality in
Eq.~(\ref{converge}) is satisfied, a more complicated OPE in terms of nonlocal
operators is still possible~\cite{Bigi:1993ex, Neubert:1993um}.

\section{Top, Higgs, and New Physics Flavor}
\addcontentsline{toc}{}{Top, Higgs, and New Physics Flavor}

\paragraph{The scale of new physics}

In the absence of direct observation of BSM particles so far, viewing the
standard model as a low energy effective theory, the search for new physics
amounts to seeking evidence for higher dimension operators invariant under the
SM gauge symmetries.

Possible dimension-6 operators include baryon and lepton number violating
operators, such as $\frac1{\Lambda^2}QQQL$.  Limits on the proton lifetime imply
$\Lambda \gsim 10^{16}\,\GeV$.  Non-SM flavor and $CP$ violation could arise
from $\frac1{\Lambda^2}Q\bar QQ\bar Q$.  The bounds on the scale of such
operators are $\Lambda \gsim 10^{4...7}\,\GeV$, depending on the generation
index of the quark fields.  Precision electroweak measurements constrain
operators of the form $\frac1{\Lambda^2}(\phi D_\mu\phi)^2$ to have $\Lambda
\gsim 10^{3...4}\,\GeV$.  These constraints are remarkable, because flavor,
$CP$, and custodial symmetry are broken by the SM itself, so it is unlikely for
new physics to have a symmetry reason to avoid introducing additional
contributions.

As mentioned earlier, there is a single type of gauge invariant dimension-5
operators made of SM fields, which give rise to neutrino masses, see
Eq.~(\ref{numass}).  The observed neutrino mass square differences hint at
scales $\Lambda > 10^{10}\,\GeV$ for these $\frac1\Lambda (L \phi)^2$ type
operators (in many models $\Lambda \sim 10^{15}\,\GeV$).  Such mass terms
violate lepton number.  It is an experimental question to determine the nature
of neutrino masses, which is what makes the search for neutrinoless double beta
decay (and determining the neutrino mass hierarchy) so important.

\paragraph{Charged lepton flavor violation (CLFV)}

The SM with vanishing neutrino masses would have predicted lepton flavor
conservation. We now know that this is not the case, hence there is no reason to
impose it on possible new physics scenarios.  In particular, if there are
TeV-scale new particles that carry lepton number (e.g., sleptons), then they
have their own mixing matrices, which could give rise to CLFV signals.  While
the one-loop SM contributions to processes such as $\mu\to e\gamma$ are
suppressed by the neutrino mass-squared differences\footnote{Estimate the $\mu
\to e \gamma$ rate in the SM.}, the NP contributions have a-priori no such
suppressions, other than the somewhat heavier scales and being generated at
one-loop in most BSM scenarios.

Within the next decade, the CLFV sensitivity will improve by about 4 orders of
magnitude, corresponding to an increase in the new physics scale probed by an
order of magnitude, possibly the largest such gain in sensitivity achievable
soon.  If any CLFV signal is discovered, we would want to measure many processes
to map out the underlying patterns, including $\mu\to e\gamma$, $\mu\to 3e$,
$\tau\to e\gamma$, $\tau\to 3e$, $\tau\to \mu\gamma$, $\tau\to 3\mu$, etc.

\paragraph{Electric dipole moments (EDM)}

The experimental bound on the neutron EDM implies that a possible dimension-4
term in the SM Lagrangian, $\theta_{\rm QCD} F\widetilde F/(16\pi^2)$, has a
coefficient $\theta_{\rm QCD} \lsim 10^{-10}$.  While there are plausible
explanations~\cite{Dine:2000cj}, we do not yet know the resolution with
certainty.  Neglecting this term, $CP$ violation in the CKM matrix only gives
rise to quark EDMs at three-loop order, and lepton EDMs at four-loop level,
resulting in EDMs below near future experimental sensitivities.  On the other
hand, new physics (e.g., supersymmetry) could generate both quark and lepton
EDMs at the one-loop level, so even if the scale of new physics is
10\,--\,100\,\TeV, observable effects could arise.

\paragraph{Top quark flavor physics}

Well before the LHC turned on, it was already certain that it was going to be a
top quark factory; the HL-LHC is expected to produce a few times $10^9$ $t\bar
t$ pairs.  In the SM, top quarks almost exclusively decay to $Wb$, as
$\big||V_{tb}| - 1 \big| \approx 10^{-3}$.  The current bounds on FCNC top
decays are at the $10^{-3}$ level, and the ultimate LHC sensitivity is expected
to reach the $10^{-5}$ to $10^{-6}$ level, depending on the decay mode.  The SM
rates are much smaller\footnote{Estimate the $t\to c Z$ and $t\to c \gamma$
branching ratios in the SM.}, so observation of any FCNC top decay signal
would be clear evidence for new physics.

There is obvious complementarity between FCNC searches in the top sector and low
energy flavor physics bounds.  Since $t_L$ is in the same $SU(2)$ doublet as
$b_L$, several operators have correlated effects in $t$ and $b$ decays.  For
some operators, mainly those involving left-handed quark fields, the low energy
constraints already exclude a detectable LHC signal, whereas other operators may
still have large enough coefficients to yield detectable effects in top FCNCs at
the LHC (see, e.g., Ref.~\cite{Fox:2007in}).

The $t\bar t$ forward-backward asymmetry provided a clear example recently of
the interplay between flavor physics and anomalies in the high energy collider
data (even those that may seem little to do with flavor at first).  The CDF
measurement in 2011, $A_{t\bar t}^{\rm FB}(m_{t\bar t} > 450\,\GeV) = 0.475 \pm
0.114$~\cite{Aaltonen:2011kc}, was stated to be $3.4\sigma$ above the NLO SM
prediction.  At the LHC, the same underlying physics would produce a rapidity
asymmetry.\footnote{Show that if in $t\bar t$ production at the Tevatron more
$t$ goes in the $p$ than in the $\bar p$~direction, then at the LHC the mean
magnitude of the $t$ quark rapidity is greater than that of the~$\bar t$.}  It
became quickly apparent that models that could account for this signal faced
severe flavor constraints.  This provides an example (with hundreds of papers in
the literature) that flavor physics will likely be crucial to understand what
the explanation of a high-$p_T$ LHC anomaly can be, and also what it cannot be. 
By now this excitement has subsided, because the significance of the Tevatron
anomaly decreased and because the LHC has not seen any anomalies in the top
production data predicted by most models (see, e.g., Ref.~\cite{Gresham:2011fx})
built to explain the Tevatron signal.

\paragraph{Higgs flavor physics}

With the discovery of a SM-like Higgs boson at the LHC, it is now clear that the
LHC is also a Higgs factory.  Understanding the properties of this particle
entails both the precision measurements of its observed (and not yet seen)
couplings predicted by the SM, and the search for possible decays forbidden in
the SM.

The source of Higgs flavor physics, obviously, is the same set of Yukawa
couplings whose structure and consequences we also seek to understand in low
energy flavor physics measurements.  While in terms of SUSY model building $m_h
\approx 125\,$GeV is challenging to understand, this mass allows experimentally
probing many Higgs production and decay channels. The fact that ultimately the
LHC will be able to probe Higgs production via (i) gluon fusion ($gg\to h$),
(ii) vector boson fusion ($q\bar q\to q\bar q h$), (iii) $W/Z$ associated
production ($q\bar q\to hZ$ or $hW$), (iv) $b/t$ associated production ($gg\to
hb\bar b$ or $ht\bar t$) sensitively depend on the Yukawa couplings and
$m_h$.\footnote{How would Higgs production and decay change if $m_t$ were, say,
50\,GeV?}

If we allow new physics to contribute to Higgs-related processes, which is
especially well motivated for loop-induced production (e.g., the dominant $gg\to
h$) and decay (e.g., $h\to\gamma\gamma$) channels, then the first evidence for
non-universal Higgs couplings to fermions was the bound on $h\to \mu^+\mu^-$
below $10\times \mbox{(SM prediction)}$, combined with the observations of $h\to
\tau^+\tau^-$ at the SM level, implicitly bounding ${\cal B}(h\to \mu^+\mu^-) /
{\cal B}(h\to \tau^+\tau^-) \lsim 0.03$.

There is an obvious interplay between the search for flavor non-diagonal Higgs
decays and indirect bounds from flavor-changing quark transitions and bounds on
CLFV in the lepton sector.  For example, $y_{e\mu} \neq 0$ would generate a
one-loop contribution to $\mu\to e \gamma$, $y_{uc} \neq 0$ would generate
$D^0$\,--\,$\D0bar$ mixing, etc.~\cite{Harnik:2012pb}.  In some cases the flavor
physics constraints imply that there is no chance to detect a particular
flavor-violating Higgs decay, while signals in some modes may be above future
direct search sensitivities.  The interplay between measurements and constraints
on flavor-diagonal and flavor-changing Higgs decay modes can provide additional
insight on which flavor models are viable (see, e.g., Ref.~\cite{Dery:2013rta}).

\paragraph{Supersymmetry and flavor}

While I hope the LHC will discover something unexpected, of the known BSM
scenarios, supersymmetry is particularly interesting, and its signals have been
worked out in great detail.  The minimal supersymmetric standard model (MSSM)
contains 44 $CP$ violating phases and 80 other $CP$ conserving flavor
parameters~\cite{Haber:1997if}.\footnote{Check this, using the counting of
couplings and broken global symmetries.}  It has long been known that flavor
physics (neutral meson mixings, $\epsilon'_K$, $\mu\to e\gamma$, $B\to
X_s\gamma$, etc.) imposes strong constraints on the SUSY parameter space.  The
MSSM also contains flavor-diagonal $CP$ violation (in addition to $\theta_{\rm
QCD}$), and the constraints from the bounds on electric dipole moments are
fairly strong on these phases if the mass scale is near 1\,TeV.

As an example, consider the $K_L$\,--\,$K_S$ mass difference.  The
squark--gluino box contribution compared to the data contains terms, roughly,
\beq\label{epsKScon}
{\Delta m_K^{\rm (SUSY)}\over \Delta m_K^{\rm (exp)}}
  \sim 10^4\, \bigg ({1\,\TeV\over\tilde m}\bigg)^{\!2}\,
  \bigg({\Delta \tilde m^2\over \tilde m^2}\bigg)^{\!2}\,
  {\rm Re} \big[(K^d_L)_{12}(K^d_R)_{12}\big],
\eeq
where $K^d_L$ ($K^d_R$) are the mixing matrices in the gluino couplings to
left-handed (right-handed) down quarks and their scalar
partners~\cite{Ligeti:2002wt}.  The constraint from $\epsilon_K$ corresponds to
replacing $10^4\, {\rm Re} \big[(K^d_L)_{12}(K^d_R)_{12}\big]$ with $10^6\, {\rm
Im} \big[(K^d_L)_{12}(K^d_R)_{12}\big]$.  The simplest supersymmetric frameworks
with parameters in the ballpark of $\tilde m = {\cal O}(1\,\TeV)$, $\Delta\tilde
m^2/\tilde m^2 = {\cal O}(0.1)$, and $(K^d_{L,R})_{ij} = {\cal O}(1)$ are
excluded by orders of magnitude.

There are several ways to address the supersymmetric flavor problems.  There are
classes of models that suppress each of the terms in Eq.~(\ref{epsKScon}):
(i)~heavy squarks, when $\tilde m \gg 1\,\TeV$ (e.g., split SUSY); (ii)
universality, when $\Delta\tilde m_{\tilde Q,\tilde D}^2 \ll \tilde m^2$ (e.g.,
gauge mediation); (iii) alignment, when $(K^d_{L,R})_{12} \ll 1$ (e.g.,
horizontal symmetry).  All viable models incorporate some of these ingredients
in order not to violate the experimental bounds.  Conversely, if SUSY is
discovered, mapping out its flavor structure will help to answer important
questions about even higher scales, e.g., the mechanism of SUSY breaking, how it
is communicated to the MSSM, etc.

A special role in constraining SUSY models is played by $D^0$\,--\,$\D0bar$
mixing, which was the first observed FCNC process in the up-quark sector.  It is
a special probe of BSM physics, because it is the only neutral meson system in
which mixing is generated by intermediate down-type quarks in the SM, or
intermediate up-type squarks in SUSY.  The constraints are thus complementary to
FCNC processes involving $K$ and $B$ mesons.  $D^0$\,--\,$\D0bar$ mixing and
FCNC in the up-quark sector are particularly important in constraining scenarios
utilizing quark-squark alignment~\cite{Nir:1993mx, Gedalia:2012pi}.

Another important implication for SUSY searches is that the LHC constraints on
squark masses are sensitive to the level of (non-)degeneracy of squarks required
to satisfy flavor constraints.  Most SUSY searches assume that the first two
generation squarks, $\tilde u_{L,R}$, $\tilde d_{L,R}$, $\tilde s_{L,R}$,
$\tilde c_{L,R}$, are all degenerate, which increases signal cross sections. 
Relaxing this assumption consistent with flavor
bounds~\cite{Gedalia:2012pi,Crivellin:2010ys}, results in substantially weaker
squark mass limits from Run~1, as low as around the 500\,GeV
scale~\cite{Mahbubani:2012qq}.

It is apparent from the above discussion that there is a tight interplay between
the implications of the non-observation of new physics at the LHC so far, and
the non-observation of deviations from the SM in flavor physics.  If there is
new physics at the TeV scale, which we hope the LHC will discover in its next
run, then we know already that its flavor structure must be rather non-generic
to suppress FCNCs, and the combination of all data will contain plenty of
additional information about the structure of new physics.  The higher the scale
of new physics, the less severe the flavor constraints are.  If NP is beyond the
reach of the LHC, flavor physics experiments may still observe robust deviations
from the SM predictions, which would point to an upper bound on the next scale
to probe.

\paragraph{Minimal flavor violation (MFV)}

The standard model without Yukawa couplings has a global $[U(3)]^5$ symmetry
($[U(3)]^3$ in the quark and $[U(3)]^2$ in the lepton sector), rotating the 3
generations of the 5 fields in Eq.~(\ref{reps}).  This is broken by the Yukawa
interactions in Eq.~(\ref{Lyuk}).  One may view the Yukawa couplings as
spurions, fields which transform under $[U(3)]^5$ in a way that makes the
Lagrangian invariant, and then the global flavor symmetry is broken by the
background values of the Yukawas.  BSM scenarios in which there are no new
sources of flavor violation beyond the Yukawa matrices are called minimal flavor
violation~\mbox{\cite{Chivukula:1987py, Hall:1990ac, D'Ambrosio:2002ex}}.  Since
the SM breaks the $[U(3)]^5$ flavor symmetry already, MFV gives a framework to
characterize ``minimal reasonable" deviations from the SM predictions.

Let us focus on the quark sector.  Under $U(3)_Q \times U(3)_u \times U(3)_d$
the transformation properties are
\beq
Q_L(3,1,1)\,,\quad u_R(1,3,1)\,,\quad d_R(1,1,3)\,,\qquad 
Y_u(3,\bar3,1)\,,\quad Y_d(3,1,\bar3)\,.
\eeq
One can choose a basis in which
\beq
Y_d = {\rm diag}(y_d\,, y_s\,, y_b)\,, \qquad
Y_u = V_{\rm CKM}^\dagger\, {\rm diag}(y_u\,, y_c\,, y_t)\,.
\eeq
To generate a flavor-changing transition, requires constructing $[U(3)]^3$
singlet terms that connect the required fields.  For example, in the down-quark
sector, the simplest terms are~\cite{D'Ambrosio:2002ex}
\beq
\bar Q_L Y_u Y_u^\dagger Q_L\,, \qquad
\bar d_R Y_d^\dagger Y_u Y_u^\dagger Q_L\,, \qquad
\bar d_R Y_d^\dagger Y_u Y_u^\dagger Y_d d_R\,.
\eeq
A useful feature of this approach is that it allows EFT-like analyses.

Consider $B\to X_s\gamma$ as an example.  We are interested in the magnitude of
a possible NP contribution to the Wilson coefficient of the operator
$\frac{X}\Lambda (\bar s_L \sigma_{\mu\nu} F^{\mu\nu} b_R)$.  A term $\bar Q_L
b_R$ is not invariant under $[U(3)]^3$.  A term $\bar Q_L Y_d\, d_R$ is
$[U(3)]^3$ invariant, but it is diagonal, so it only connects same generation
fields.  The first non-vanishing contribution comes from $\bar Q_L Y_u
Y_u^\dagger Y_d\, d_R$, which has a $V_{tb}V_{ts}^*\, y_t^2 y_b (\bar s_L  b_R)$
component.  We learn that in MFV models, in general, $X \propto y_b V_{tb}
V_{ts}^*$, as is the case in the SM.

Thus, in MFV models, most flavor-changing operators ``automatically" have their
SM-like suppressions, proportional to the same CKM elements, quark masses from
chirality flips, etc.  Therefore, the scale of MFV models can be ${\cal
O}(1\,\TeV)$ without violating flavor physics bounds, thus solving the new
physics flavor puzzle.  Originally introduced for technicolor
models~\cite{Chivukula:1987py}, gauge-mediated supersymmetry breaking provides
another well known scenario in which MFV is expected to be a good approximation.

MFV models have important implications for new particle searches, too.  Since
the only quark flavor-changing parameters are the CKM elements, and the ones
that couple the third generation to the lighter ones are very small, in MFV
models new particles that decay to a single final quark (and other particles)
decay to either a third generation quark or to quarks from the first two
generations, but (to a good approximation) not to both~\cite{Grossman:2007bd}.

The MFV ansatz can be incorporated into models that do not contain explicitly
flavor breaking unrelated to Yukawa couplings.  MFV is not expected to be an
exact symmetry, but it may be a useful organizing principle to understand
details of the new physics we soon hope to get a glimpse of.

\section{Summary}
\addcontentsline{toc}{}{Summary}

An essential feature of flavor physics is its ability to probe very high scales,
beyond the masses of particles that can be produced on-shell in colliders. 
Flavor physics can also teach us about properties of TeV-scale new physics,
that cannot be learned from the direct production of new particles.  

Some of the main points I tried to explain in these lectures were:

\begin{itemize}

\item Flavor-changing neutral currents and meson mixing probe scales well above
the masses of particles colliders can produce, and provide strong constraints on
TeV-scale new physics.

\item $CP$ violation is always the result of interference
phenomena, without a classical analog.

\item The KM phase has been established as the dominant source of $CP$ violation
in flavor-changing processes.

\item Tremendous progress will continue: Until $\sim$\,10 years ago, more than
${\cal O}(1)$ deviations from the SM were possible; at present ${\cal O}(20\%)$
corrections to most FCNC processes are still allowed; in the future,
sensitivities of a few percent will be reached. 

\item The future goal is not measuring SM parameters better, but to search for
corrections to the SM, and to learn about NP as much as possible.

\item Direct information on new particles and their influence on flavor-changing
processes will both be crucial to understand the underlying physics.

\item The sensitivity of future experiments in a number of important processes
is only limited by statistics, not theory.

\item The interesting (and fun) interplay between theoretical and experimental
developments in flavor physics will continue.

\end{itemize}

\begin{figure}[t]
\centerline{\includegraphics*[height=6cm]{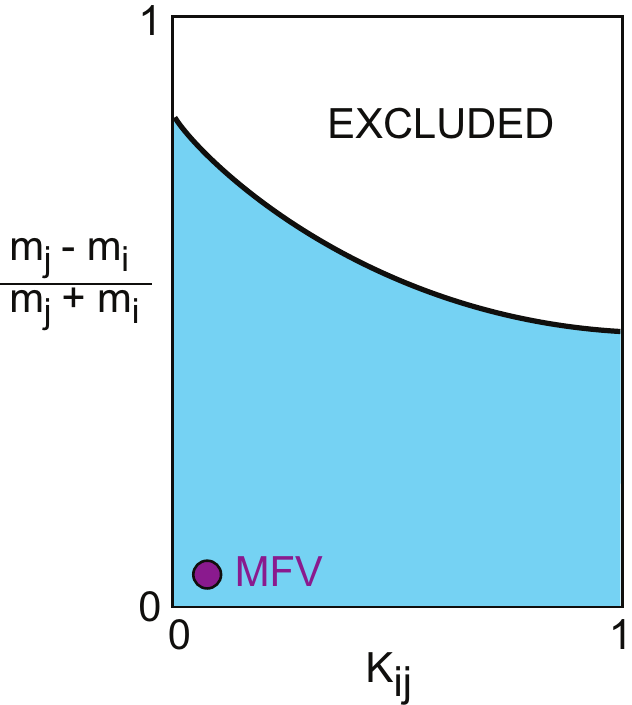} \hfill
\includegraphics*[height=6cm]{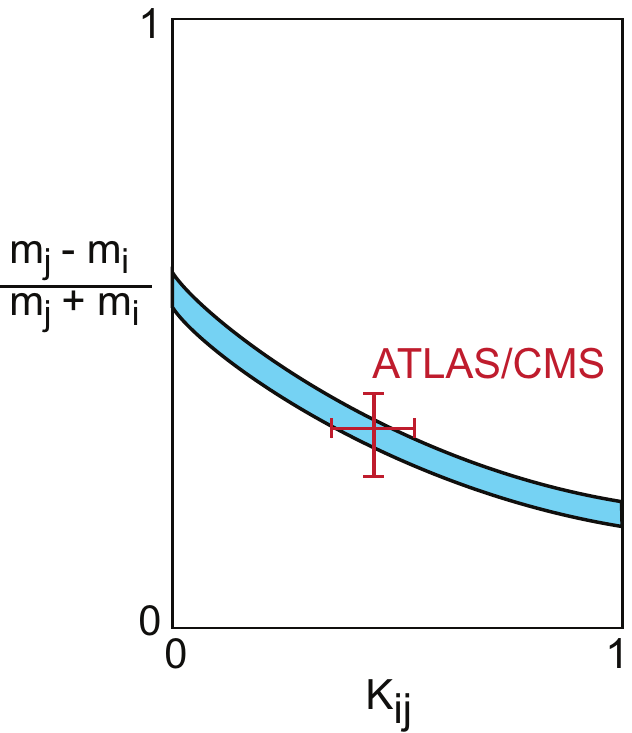}}
\caption{Schematic description of the constraints on the mass splitting,
$(m_i-m_j)/(m_i+m_j)$, and mixing angle, $K_{ij}$, between squarks (or
sleptons).  Left: typical constraint from not observing deviations from the SM. 
The fact that ${\cal O}(1)$ splittings and mixings are excluded constitutes the
new physics flavor puzzle.  Right: possible future scenario where ATLAS/CMS
measurements fit flavor physics signals of NP.  (From
Ref.~\cite{Grossman:2009dw}.)}
\label{fig:future}
\end{figure}

At present, both direct production and flavor physics experiments only give
bounds on new physics.  The constraints imply that if new physics is accessible
at the LHC, it is likely to have flavor suppression factors similar to the SM. 
In many models (e.g., the MSSM), measurements or bounds on FCNC transitions
constrain the product of certain mass splittings times mixing parameters divided
by the square of the new physics scale.  If the LHC discovers new physics, then
in principle the mass splittings and mixing parameters can be measured
separately.  If flavor physics experiments establish a deviation from the SM in
a related process, the combination of LHC and flavor data can be very powerful
to discriminate between models.  The consistency of measurements could
ultimately tell us that we understand the flavor structure of new physics and
how the new physics flavor puzzle is solved.  The present situation and an
(optimistic) future scenario for supersymmetry are shown in
Fig.~\ref{fig:future}.  Let's hope that we shall have the privilege to think
about such questions, motivated by data, in the coming~years.

\section*{Acknowledgments}

I thank Lance Dixon and Frank Petriello for surprisingly successful
arm-twisting (and patience), so that these notes got written up, and Marat
Freytsis, Yonit Hochberg, and Dean Robinson for helpful comments.
This work was supported by the Office of Science, Office of High Energy
Physics, of the U.S.\ Department of Energy under contract DE-AC02-05CH11231.

\end{document}